\documentclass[11pt]{article}

\usepackage[margin=1in]{geometry}
\usepackage{amsmath}
\usepackage{amssymb,amsfonts}
\usepackage{mathtools}
\numberwithin{equation}{section}
\usepackage{placeins}
\usepackage{bm}
\usepackage{booktabs}
\usepackage{array}
\usepackage{makecell}
\usepackage{longtable}
\usepackage{graphicx}
\usepackage{caption}
\usepackage{subcaption}
\usepackage{pgfplots}
\usepackage{hyperref}
\hypersetup{hidelinks}
\usepackage{float}
\usepackage{setspace}
\usepackage{titlesec}
\usepackage{lmodern}
\usepackage[expansion=false]{microtype}
\usepackage{physics}
\usepackage{enumitem}
\usepackage{xcolor}
\usepackage{dsfont} 

\usepackage{sidecap}
\usepackage{eso-pic}

\newcommand{\iu}{\mathrm{i}}

\usepackage{scalerel}[2016/12/29]

\allowdisplaybreaks
\pgfplotsset{compat=1.18}

\makeatletter
\newcommand{\appsection}[1]{%
  \clearpage
  \refstepcounter{section}%
  \phantomsection%
  \section*{Appendix \thesection\space #1}%
  \addcontentsline{toc}{section}{Appendix \thesection\space #1}%
  \setcounter{subsection}{0}%
  \setcounter{subsubsection}{0}%
  \setcounter{equation}{0}%
}
\makeatother

\title{\Large{Group Quantization and Mellin Representations of the Heston Model}}

\author{
Santiago Garc\'ia\\
Wells Fargo Securities\\
\texttt{santiago.garcia22@gmail.com}
}

\begin{document}

\maketitle

\begin{abstract}
We formulate the Heston stochastic-volatility model within an
\emph{Affine Holonomy Group Quantization} (AHGQ) framework that extends
Group Quantization to affine models with state-dependent covariance. The
Heston pricing symbol is decomposed into a homogeneous quadratic sector,
described by a centrally extended symplectic Lie group, and a complementary
sector represented through multiplicative holonomy. Their combination
determines the Heston Poincar\'e--Cartan form and its characteristic
field. In the momentum representation, spatial polarization reduces this
field to the affine Riccati system underlying the Heston transform
solution. Mellin decomposition then leads to a projective representation of
the scalar Riccati equation and to a European option-pricing formula. The
resulting representation is compared numerically with the standard
semi-analytic Heston formula, and the Black--Scholes reduction is recovered
as a limiting case. The construction provides a common geometric description
of symplectic transport, affine Riccati dynamics, holonomy, and transform
pricing.
\end{abstract}

\newpage
\tableofcontents
\newpage

\section{Introduction}

The Heston stochastic-volatility model~\cite{Heston1993} is one of the
canonical affine models of mathematical finance. Its analytical tractability follows
from the affine structure of its pricing generator: transform methods reduce
the pricing problem to exponential-affine kernels whose coefficients satisfy
Riccati equations. In the standard affine-transform approach
\cite{DuffiePanSingleton2000,DuffieFilipovicSchachermayer2003}, these equations
arise from the exponential-affine ansatz and the affine form of the generator.
The purpose of this paper is to describe the same structure through
Group Quantization and the associated holonomy geometry.

The construction extends the Group Quantization approach developed in
\cite{quadratic} for quadratic financial diffusion models. In the Heston
model, however, the covariance depends on the instantaneous variance \(v\),
so an additional geometric ingredient is required. We therefore decompose
the Heston affine pricing symbol according to the geometric role of its
terms into a homogeneous quadratic sector \(C_s\) and a complementary
sector \(C_H\), the latter represented through holonomy. We refer to the
resulting framework as \emph{Affine Holonomy Group Quantization} (AHGQ).

The homogeneous quadratic sector generates a four-dimensional linear
symplectic transport \(M_s(t)\). This transport determines the corresponding
finite Lie-group structure. The symplectic form, together with \(M_s(t)\),
induces a two-cocycle that defines a multiplicative
\(\mathbb R_+\)-central extension of this group. The positive central
coordinate plays the role of a pricing scale rather than a
quantum-mechanical phase. The standard Group Quantization construction can
then be applied to the centrally extended symplectic group, yielding its
invariant generators and Cartan form.

The remaining affine terms are not incorporated into the finite Lie-group
multiplication. Instead, \(C_H\) defines a multiplicative holonomy along
paths in \(G^s\), with a groupoid organizing their composition and the
holonomy acting only on the positive central fiber. Lie-group multiplication
and path composition therefore remain distinct structures.

Adding the holonomy contribution to the finite-sector Cartan form gives the
full Heston Poincar\'e--Cartan form, whose characteristic module determines
the Heston characteristic field. Spatial polarizations are chosen from the spatial generators of the finite
symplectic Lie-group sector only to select the representation variables. The quadratic-sector time generator is therefore not imposed as
a spatial polarization operator; the full characteristic direction is
treated separately through \(X_\Theta\).

Pricing is carried out primarily in the momentum representation. After
spatial polarization, \(X_\Theta\) reduces to a first-order momentum operator
containing the nonzero component of the vector-valued Riccati symbol and the
state-independent affine term. Mellin transformation fixes the conserved
log-price momentum and reduces the remaining dynamics to the scalar Heston
Riccati equation. The complementary coordinate polarization yields a
coordinate realization of the Heston pricing operator.

The paper is organized in two stages. Sections~\ref{sec:gaq-background}--\ref{sec:heston-cartan-characteristics}
develop the geometric construction, from the affine-symbol decomposition to
the full Poincar\'e--Cartan form and characteristic field. Section~\ref{sec:heston-spatial-polarization}
constructs the momentum and coordinate representations, and
Section~\ref{sec:financial-applications} develops Mellin pricing, the
projective Riccati representation, and numerical validation. The
Black--Scholes reduction is treated separately in
Appendix~\ref{app:black-scholes-reduction}.

\section{Notation for the Heston Holonomy Construction}
\label{sec:notation-holonomy}

\begin{table}[H]
\centering
\renewcommand{\arraystretch}{1.2}
\begin{tabular}{ll}
\toprule
Symbol & Meaning \\
\midrule
\(\mathbf x=(x,v)^T\) & Heston state vector \\
\(\mathbf p=(p_x,p_v)^T\) & conjugate momentum vector \\
\(\mathbf a=(\mathbf x,\mathbf p)^T\) & four-dimensional phase-space vector \\
\(C_A\) & Heston affine pricing symbol \\
\(F(\mathbf p)\) & state-independent part of the Heston affine symbol \\
\(\mathbf R(\mathbf p)\) & vector-valued affine Riccati symbol \\
\(R(p_x,p_v)\) & nonzero variance component of \(\mathbf R\) \\
\(C_s\) & homogeneous quadratic Heston sector \\
\(C_H\) & complementary affine sector represented through holonomy \\
\(K_s\) & generator of the linear Heston symplectic transport \\
\(M_s(t)=e^{tK_s}\) & four-dimensional linear symplectic transport \\
\(G^s\) & Lie group associated with \(M_s(t)\) \\
\(\widetilde G^s\) & \(\mathbb R_+\)-central extension of \(G^s\) \\
\(\mathcal G\rightrightarrows G^s\) & thin-path groupoid of the finite symplectic sector \\
\(\mathcal H[\gamma]\) & multiplicative holonomy associated with \(C_H\) \\
\(\Xi=\zeta\partial_\zeta\) & positive central generator \\
\(\Theta\) & full Heston Poincar\'e--Cartan form \\
\(X_\Theta\) & characteristic field of \(\Theta\) \\
\(D(\tau,q)\) & transported variance momentum \\
\(A(\tau,q)\) & affine amplitude along the momentum characteristic \\
\(q\) & Mellin spectral parameter \\
\(\tau\) & time to maturity in the pricing section \\
\bottomrule
\end{tabular}
\caption{Notation used in the Heston symplectic-group and holonomy construction.}
\label{tab:heston-holonomy-notation}
\end{table}

\begin{table}[h!]
\centering
\small
\begin{tabular}{@{}p{0.34\textwidth}p{0.60\textwidth}@{}}
\toprule
\textbf{Geometric object} & \textbf{Financial meaning} \\
\midrule
\(\mathbb R^+\)-central extension
& positive pricing and discounting scale \\

\(C_A\)
& Heston affine pricing symbol \\

\(C_s\)
& homogeneous quadratic sector generating finite symplectic transport \\

\(C_H\)
& complementary affine sector represented through holonomy \\

\(\mathcal P_{\mathrm{coord}}^s\)
& spatial coordinate polarization \\

\(\mathcal L_A\)
& coordinate-space realization of the Heston pricing operator \\

\(\mathcal P_{\mathrm{mom}}^s\)
& spatial momentum polarization \\

Momentum-polarized functions
& momentum-space representation of European instruments \\

\(X_\Theta^{\mathrm{mom}}\)
& reduced momentum characteristic operator \\

\(\mathbf R(\mathbf p)\)
& vector-valued affine Riccati symbol \\

\(R(p_x,p_v)\)
& nonzero variance component of \(\mathbf R\) \\

\(D(\tau,q)\)
& transported variance momentum \\

\(G(\tau;q,q')\)
& momentum propagator for the reduced Mellin components \\

\(A(\tau,q)\)
& diagonal amplitude of the momentum propagator \\

\(\mathcal K_q(\tau,x,v)\)
& propagated Mellin mode used in the numerical pricing formula \\

\(M_R(\tau,q)\)
& auxiliary linear transport of the projective Riccati flow \\
\bottomrule
\end{tabular}
\caption{Dictionary between the geometric Heston construction and the corresponding financial objects.}
\label{tab:heston-geometric-financial-dictionary}
\end{table}
\section{Geometric and Group Quantization Background}
\label{sec:gaq-background}

In this paper, we use only a few elements of the Group Approach to
Quantization (GAQ) developed by Aldaya and de Azc\'arraga
\cite{aldaya0,wolf11,AldayaGarcia,aldaya22,aldaya6}. The dynamics is encoded
in differential operators derived from an underlying group structure. Two
compatible families of operators arise: one is used to impose constraints
that select the variables retained in a representation, while the other acts
on the resulting functions.

GAQ also introduces an additional multiplicative variable through a central
extension. In its original quantum-mechanical formulation, GAQ usually takes
\(U(1)\) as the central group, so that this variable is interpreted as a
phase. Here we take \(\mathbb R_+\), so that the central coordinate acts
instead as a positive pricing or discounting scale. We write
\begin{equation}
\Xi=\zeta\partial_\zeta,\qquad \zeta\in\mathbb R_+.
\label{eq:heston-central-generator}
\end{equation}

A polarization specifies which variables are retained in a representation.
For the construction below, the momentum polarization retains the conjugate
variables and leads to the Heston Riccati representation, while the
coordinate polarization retains the state variables and gives the
complementary coordinate representation.

The standard GAQ construction follows a definite sequence. One first
specifies the group law and its central extension. The corresponding left-
and right-invariant generators are then obtained by differentiation. The
central extension determines a distinguished Cartan form and its
characteristic subalgebra. A polarization is chosen from the left-invariant
generators to select a representation, while the right-invariant generators
descend to operators acting in that representation.

Other standard aspects of GAQ, such as the construction of Noether invariants
and an associated Lagrangian formulation, are not used in the present
development. We restrict attention to the group structure, central extension,
invariant generators, Cartan form, characteristic module, and polarizations
required for the pricing representations considered below.

In Heston, the finite-group GAQ procedure is applied to the homogeneous
quadratic sector \(C_s\), while the complementary sector \(C_H\) enters through
multiplicative holonomy. The following sections construct these ingredients
explicitly.


\section{Heston Affine Holonomy Group Quantization}

We use the term \emph{Affine Holonomy Group Quantization} (AHGQ) for the
framework developed below for the Heston model. This section constructs its
model-specific ingredients.

\subsection{Heston Affine Pricing Symbol}
\label{sec:heston-affine-symbol}

Let
\begin{equation*}
x=\log S,
\qquad
v=\text{instantaneous variance},
\end{equation*}
and introduce the state and momentum vectors
\begin{equation}
\mathbf x=(x,v)^T,
\qquad
\mathbf p=(p_x,p_v)^T.
\label{eq:heston-state-momentum-vectors}
\end{equation}

\begin{table}[ht]
\centering
\begin{tabular}{ll}
\toprule
Symbol & Meaning \\
\midrule
\(r\) & Risk-free rate \\
\(\delta\) & Dividend yield \\
\(\kappa\) & Mean-reversion speed of variance \\
\(\theta\) & Long-run variance level \\
\(\sigma_\nu\) & Volatility of variance \\
\(\rho\) & Correlation between stock and variance shocks \\
\(\mathbf b\) & Constant drift vector \\
\(\mathsf B\) & Linear state-drift matrix \\
\(\mathsf A_v\) & Coefficient matrix of the variance-dependent covariance \\
\bottomrule
\end{tabular}
\caption{Heston model parameters and affine data used below.}
\label{tab:heston-symbols}
\end{table}

The Heston affine pricing symbol may be written in matrix form as
\begin{equation}
C_A(\mathbf x,\mathbf p)
=
\mathbf b^T\mathbf p
+\mathbf x^T\mathsf B^T\mathbf p
+\frac12v\,\mathbf p^T\mathsf A_v\mathbf p
-r.
\label{eq:heston-full-affine-symbol}
\end{equation}

The affine drift data are
\begin{equation}
\mathbf b=(r-\delta,\kappa\theta)^T,
\qquad
\mathsf B=
\begin{pmatrix}
0&-\frac12\\
0&-\kappa
\end{pmatrix}.
\label{eq:heston-affine-drift-data}
\end{equation}

The Heston covariance has no state-independent component. Its
state-dependent covariance matrix is \(v\mathsf A_v\), where
\begin{equation}
\mathsf A_v
=
\begin{pmatrix}
1&\rho\sigma_\nu\\
\rho\sigma_\nu&\sigma_\nu^2
\end{pmatrix}.
\label{eq:heston-Av}
\end{equation}

In coordinates,
\begin{equation}
\begin{aligned}
C_A
&=
(r-\delta)p_x+\kappa\theta p_v-r
-v\left(\frac12p_x+\kappa p_v\right)
\\
&\qquad
+\frac12v\left(
p_x^2+2\rho\sigma_\nu p_xp_v+\sigma_\nu^2p_v^2
\right).
\end{aligned}
\label{eq:heston-affine-symbol-coordinates}
\end{equation}

The affine dependence on the state variables can equivalently be written as
\begin{equation}
C_A(\mathbf x,\mathbf p)
=
F(\mathbf p)+\mathbf x^T\mathbf R(\mathbf p),
\label{eq:heston-affine-form}
\end{equation}
where the state-independent part is
\begin{equation}
F(\mathbf p)
=
\mathbf b^T\mathbf p-r
=
(r-\delta)p_x+\kappa\theta p_v-r,
\label{eq:heston-F-holonomy}
\end{equation}
and the vector-valued Riccati symbol is
\begin{equation}
\mathbf R(\mathbf p)
=
\mathsf B^T\mathbf p
+
\frac12
\begin{pmatrix}
0\\
\mathbf p^T\mathsf A_v\mathbf p
\end{pmatrix}.
\label{eq:heston-R-definition}
\end{equation}

For Heston, the first component of \(\mathbf R\) vanishes, so its only nonzero component is \(R(p_x,p_v)\).
\begin{equation}
\mathbf R(\mathbf p)
=
\begin{pmatrix}
0\\
R(p_x,p_v)
\end{pmatrix},
\qquad
R(p_x,p_v)
=
\frac12(p_x^2-p_x)
+(\rho\sigma_\nu p_x-\kappa)p_v
+\frac12\sigma_\nu^2p_v^2.
\label{eq:heston-R-holonomy}
\end{equation}

Since \(\mathbf R\) depends only on the momentum variables, the affine form
\eqref{eq:heston-affine-form} also implies
\begin{equation}
\nabla_{\mathbf x}C_A=\mathbf R.
\label{eq:heston-R-gradient}
\end{equation}

For the AHGQ construction, the same affine symbol is decomposed according to the
geometric role of its terms:
\begin{equation}
C_A=C_s+C_H.
\label{eq:heston-symbol-split}
\end{equation}

The homogeneous quadratic sector is
\begin{equation}
C_s
=
\mathbf x^T\mathsf B^T\mathbf p
=
-v\left(\frac12p_x+\kappa p_v\right),
\label{eq:heston-Cs-holonomy}
\end{equation}
while the complementary sector is
\begin{equation}
C_H
=
\mathbf b^T\mathbf p-r
+\frac12v\,\mathbf p^T\mathsf A_v\mathbf p
=
(r-\delta)p_x+\kappa\theta p_v-r
+\frac12v\left(
p_x^2+2\rho\sigma_\nu p_xp_v+\sigma_\nu^2p_v^2
\right).
\label{eq:heston-CH-holonomy}
\end{equation}

The sector \(C_s\) generates the finite-dimensional symplectic transport,
while \(C_H\) gives the complementary contribution represented through holonomy.


\subsection{Linear Symplectic Transport Associated with \texorpdfstring{\(C_s\)}{Cs}}
\label{sec:heston-linear-symplectic-transport}

Let
\begin{equation*}
\mathbf a=(\mathbf x,\mathbf p)^T=(x,v,p_x,p_v)^T,
\qquad
J_4=
\begin{pmatrix}
0&I_2\\
-I_2&0
\end{pmatrix}.
\end{equation*}

The homogeneous quadratic sector in
\eqref{eq:heston-Cs-holonomy} can be written as
\begin{equation}
C_s=\frac12\mathbf a^TJ_4K_s\mathbf a,
\label{eq:heston-Cs-symplectic-form}
\end{equation}
where
\begin{equation}
K_s=
\begin{pmatrix}
-\mathsf B&0\\
0&\mathsf B^T
\end{pmatrix}
=
\begin{pmatrix}
0&\frac12&0&0\\
0&\kappa&0&0\\
0&0&0&0\\
0&0&-\frac12&-\kappa
\end{pmatrix}.
\label{eq:heston-Ks}
\end{equation}

Since
\begin{equation*}
K_s^TJ_4+J_4K_s=0,
\end{equation*}
the matrix exponential generated by \(K_s\) defines the symplectic transport
\begin{equation}
M_s(t)=e^{tK_s}\in Sp(4,\mathbb R),\qquad
M_s(t'+t)=M_s(t')M_s(t),\qquad
M_s(t)^TJ_4M_s(t)=J_4.
\label{eq:heston-Ms}
\end{equation}

For \(\kappa\neq0\), define
\begin{equation*}
a_t:=\frac{e^{\kappa t}-1}{2\kappa},
\qquad
b_t:=\frac{1-e^{-\kappa t}}{2\kappa}.
\end{equation*}
Then the symplectic transport takes the explicit form
\begin{equation}
M_s(t)
=
\begin{pmatrix}
1&a_t&0&0\\
0&e^{\kappa t}&0&0\\
0&0&1&0\\
0&0&-b_t&e^{-\kappa t}
\end{pmatrix}
=
\begin{pmatrix}
e^{-t\mathsf B}&0\\
0&e^{t\mathsf B^T}
\end{pmatrix}.
\label{eq:heston-Ms-explicit}
\end{equation}
Equivalently,
\begin{equation*}
e^{-t\mathsf B}
=
\begin{pmatrix}
1&a_t\\
0&e^{\kappa t}
\end{pmatrix},
\qquad
e^{t\mathsf B^T}
=
\begin{pmatrix}
1&0\\
-b_t&e^{-\kappa t}
\end{pmatrix}.
\end{equation*}

By construction, \(M_s(t)\) is determined entirely by the homogeneous
quadratic sector \(C_s\). The state-dependent covariance terms containing
\(\rho\) and \(\sigma_\nu\) are not part of this finite linear symplectic
transport; they remain in \(C_H\) and enter through the holonomy below.


\subsection{Lie Group Associated with the Heston Symplectic Transport}
\label{sec:heston-symplectic-group}

Define the Lie group \(G^s\) as the semidirect product of time translations
with phase-space translations transported by \(M_s(t)\):
\begin{equation}
G^s:=\mathbb R\ltimes_{M_s}\mathbb R^4.
\label{eq:heston-Gs}
\end{equation}

An element is written \(g=(t,\mathbf a)\). The multiplication is
\begin{equation}
(t',\mathbf a')\star^s(t,\mathbf a)
=
\left(t'+t,\,M_s(t)\mathbf a'+\mathbf a\right).
\label{eq:heston-symplectic-group-law}
\end{equation}

Associativity follows from the group property of \(M_s(t)\). We use the
convention in which the time coordinate of the second factor transports the
phase-space coordinate of the first factor. The unit and inverse are
\begin{equation*}
\mathbf 1=(0,\mathbf 0),\qquad
(t,\mathbf a)^{-1}=\bigl(-t,-M_s(-t)\mathbf a\bigr).
\end{equation*}

The symplectic transport determines the two-cocycle
\begin{equation}
\epsilon_2^s(g',g)
=
\frac12
\bigl(M_s(t)\mathbf a'\bigr)^T J_4\mathbf a
=
\frac12
\mathbf a'^TM_s(t)^TJ_4\mathbf a.
\label{eq:heston-symplectic-cocycle}
\end{equation}

Because \(M_s(t)\) is a one-parameter symplectic group,
\(\epsilon_2^s\) satisfies the normalized two-cocycle identity.
It therefore defines a central extension by the multiplicative group
\(\mathbb R_+\):
\begin{equation}
\widetilde G^s=G^s\times\mathbb R_+,
\label{eq:heston-central-extension}
\end{equation}
where \(\zeta\in\mathbb R_+\) denotes the positive central coordinate.

An element of the extended group \(\widetilde G^s\) is written \((g,\zeta)\). The
corresponding multiplication law is
\begin{equation}
(g',\zeta')\widetilde\star^s(g,\zeta)
=
\left(
g'\star^s g,\,
\zeta'\zeta
\exp\!\bigl(\epsilon_2^s(g',g)\bigr)
\right).
\label{eq:heston-central-group-law}
\end{equation}


\subsection{Thin-Path Groupoid and Heston Holonomy}
\label{sec:heston-holonomy-action}

The complementary sector \(C_H\) is represented by a path-dependent
multiplicative holonomy.

Let
\begin{equation}
\mathcal G\rightrightarrows G^s
\label{eq:heston-thin-path-groupoid}
\end{equation}
denote the thin-path groupoid of \(G^s\); see, for example,
\cite{Meneses2021}. An arrow
\begin{equation}
[\gamma]:g_0\longrightarrow g_1,
\qquad
g_0,g_1\in G^s,
\label{eq:heston-thin-path-arrow}
\end{equation}
is represented by a path in \(G^s\), and composable arrows are combined by
path concatenation.

The complementary sector defines the multiplicative holonomy
\begin{equation}
\mathcal H[\gamma]
=
\exp\!\left(
-\int_\gamma C_H\,dt
\right).
\label{eq:heston-affine-holonomy}
\end{equation}
For composable paths,
\begin{equation}
\mathcal H[\gamma_2\circ\gamma_1]
=
\mathcal H[\gamma_2]\mathcal H[\gamma_1].
\label{eq:heston-holonomy-composition}
\end{equation}

The holonomy acts only on the positive central fiber of
\(\widetilde G^s\to G^s\). For an arrow
\([\gamma]:g_0\to g_1\),
\begin{equation}
(g_0,\zeta)
\longmapsto
\bigl(g_1,\zeta\,\mathcal H[\gamma]\bigr).
\label{eq:heston-holonomy-fiber-action}
\end{equation}

Thus the groupoid arrow transports the base point from \(g_0\) to \(g_1\),
while \(\mathcal H[\gamma]\) gives the corresponding multiplicative change
in the pricing scale. This path composition is separate from the Lie-group
multiplication of \(G^s\).
\section{Invariant Generators of the Finite Symplectic Lie-Group Sector}
\label{sec:heston-invariant-generators}

The invariant vector fields are the infinitesimal form of the composition law
of the centrally extended Lie group \(\widetilde G^s\). In the GAQ framework,
a subset of the left-invariant fields defines the constraints that select a
representation, while the right-invariant fields provide operators acting on
the resulting functions.

Derivations of the finite central extension and the left-invariant fields are
collected in Appendix~\ref{app:invariant-cartan-geometry}.

We write
\begin{equation*}
\mathbf a=(\mathbf x,\mathbf p)^T=(x,v,p_x,p_v)^T,\qquad
\nabla_{\mathbf a}=(\nabla_{\mathbf x},\nabla_{\mathbf p})^T,\qquad
\Xi=\zeta \, \partial_\zeta.
\end{equation*}

\subsection{Left-Invariant Generators}
\label{sec:heston-left-invariant}

The left-invariant generators are obtained by differentiating the group
composition law with respect to the unprimed coordinates and evaluating at
the identity. In symplectic form,
\begin{equation}
L_t^s=\partial_t+(K_s\mathbf a)^T\nabla_{\mathbf a},
\qquad
\boldsymbol L_{\mathbf a}^{\,s}
=
\nabla_{\mathbf a}-\frac12J_4\mathbf a\,\Xi,
\qquad
L_\zeta^s=\Xi.
\label{eq:heston-left-invariant-generators}
\end{equation}

The symplectic flow generated by \(C_s\) enters the left-invariant time
generator. Using
\begin{equation*}
K_s=
\begin{pmatrix}
-\mathsf B&0\\
0&\mathsf B^T
\end{pmatrix},
\qquad
J_4=
\begin{pmatrix}
0&I_2\\
-I_2&0
\end{pmatrix},
\end{equation*}
the same fields take the phase-space form
\begin{equation}
\begin{aligned}
L_t^s
&=
\partial_t
-(\mathsf B\mathbf x)^T\nabla_{\mathbf x}
+(\mathsf B^T\mathbf p)^T\nabla_{\mathbf p},
&\qquad
L_\zeta^s
&=
\Xi,
\\
\boldsymbol L_{\mathbf x}^{\,s}
&=
\nabla_{\mathbf x}-\frac12\mathbf p\,\Xi,
&\qquad
\boldsymbol L_{\mathbf p}^{\,s}
&=
\nabla_{\mathbf p}+\frac12\mathbf x\,\Xi.
\end{aligned}
\label{eq:heston-left-invariant-phase-form}
\end{equation}

For the Heston matrix \(\mathsf B\), these fields are
\begin{equation}
\begin{aligned}
L_t^s
&=
\partial_t+\frac12v\,\partial_x+\kappa v\,\partial_v
-\left(\frac12p_x+\kappa p_v\right)\partial_{p_v},
&\qquad
L_\zeta^s
&=
\Xi,
\\
L_x^s
&=
\partial_x-\frac12p_x\Xi,
&\qquad
L_v^s
&=
\partial_v-\frac12p_v\Xi,
\\
L_{p_x}^s
&=
\partial_{p_x}+\frac12x\Xi,
&\qquad
L_{p_v}^s
&=
\partial_{p_v}+\frac12v\Xi.
\end{aligned}
\label{eq:heston-left-components}
\end{equation}

\subsection{Right-Invariant Generators}
\label{sec:heston-right-invariant}

The right-invariant generators are obtained by differentiating the group
composition law with respect to the primed coordinates and evaluating at the
identity. In symplectic form,
\begin{equation}
R_t^s=\partial_t,
\qquad
\boldsymbol R_{\mathbf a}^{\,s}
=
M_s(t)^T
\left(
\nabla_{\mathbf a}+\frac12J_4\mathbf a\,\Xi
\right),
\qquad
R_\zeta^s=\Xi.
\label{eq:heston-right-invariant-generators}
\end{equation}
The transport \(M_s(t)\) appears in the right-invariant phase-space
generators. Using \eqref{eq:heston-Ms-explicit}, the fields are

\begin{equation}
\begin{aligned}
R_t^s
&=
\partial_t,
&\qquad
R_\zeta^s
&=
\Xi,
\\
\boldsymbol R_{\mathbf x}^{\,s}
&=
e^{-t\mathsf B^T}
\left(
\nabla_{\mathbf x}+\frac12\mathbf p\,\Xi
\right),
&\qquad
\boldsymbol R_{\mathbf p}^{\,s}
&=
e^{t\mathsf B}
\left(
\nabla_{\mathbf p}-\frac12\mathbf x\,\Xi
\right).
\end{aligned}
\label{eq:heston-right-invariant-phase-form}
\end{equation}

Using the block exponentials following \eqref{eq:heston-Ms-explicit},
the right-invariant fields can also be expressed as
\begin{equation}
\begin{aligned}
R_t^s
&=
\partial_t,
&\qquad
R_\zeta^s
&=
\Xi,
\\
R_x^s
&=
\partial_x+\frac12p_x\Xi,
&\qquad
R_v^s
&=
a_t\left(\partial_x+\frac12p_x\Xi\right)
+e^{\kappa t}\left(\partial_v+\frac12p_v\Xi\right),
\\
R_{p_x}^s
&=
\partial_{p_x}-b_t\partial_{p_v}
-\frac12\left(x-b_t v\right)\Xi,
&\qquad
R_{p_v}^s
&=
e^{-\kappa t}
\left(\partial_{p_v}-\frac12v\Xi\right).
\end{aligned}
\label{eq:heston-right-components}
\end{equation}

\subsection{Commutators}
\label{sec:heston-invariant-commutators}

Let \(L_{a_i}^s\) denote the components of
\(\boldsymbol L_{\mathbf a}^{\,s}\). In symplectic form, the nonzero
left-invariant structural commutators are encoded by
\begin{equation}
[L_{a_i}^s,L_{a_j}^s]=(J_4)_{ij}\Xi,
\qquad
[L_t^s,\boldsymbol L_{\mathbf a}^{\,s}]
=
-K_s^T\boldsymbol L_{\mathbf a}^{\,s}.
\label{eq:heston-left-commutators-symplectic}
\end{equation}

Using
\begin{equation*}
K_s=
\begin{pmatrix}
-\mathsf B&0\\
0&\mathsf B^T
\end{pmatrix},
\qquad
J_4=
\begin{pmatrix}
0&I_2\\
-I_2&0
\end{pmatrix},
\end{equation*}
the same relations take the phase-space form
\begin{equation}
\begin{aligned}
[L_{x_i}^s,L_{p_i}^s]
&=\Xi,\qquad i=1,2,
&\qquad
[L_t^s,\boldsymbol L_{\mathbf x}^{\,s}]
&=\mathsf B^T\boldsymbol L_{\mathbf x}^{\,s},
\\
[L_t^s,\boldsymbol L_{\mathbf p}^{\,s}]
&=-\mathsf B\boldsymbol L_{\mathbf p}^{\,s}.
\end{aligned}
\label{eq:heston-left-commutators-phase}
\end{equation}

For the Heston matrix \(\mathsf B\), the nonzero commutators are
\begin{equation}
\begin{aligned}
[L_x^s,L_{p_x}^s]
&=\Xi,
&\qquad
[L_v^s,L_{p_v}^s]
&=\Xi,
\\
[L_t^s,L_v^s]
&=-\frac12L_x^s-\kappa L_v^s,
&\qquad
[L_t^s,L_{p_x}^s]
&=\frac12L_{p_v}^s,
\\
[L_t^s,L_{p_v}^s]
&=\kappa L_{p_v}^s.
\end{aligned}
\label{eq:heston-left-commutators-components}
\end{equation}

All other left-invariant commutators vanish. The right-invariant fields
satisfy the corresponding commutation relations with an overall minus sign.

These commutation relations are independent of the variance coordinate
\(v\) and of the covariance parameters \(\rho\) and \(\sigma_\nu\). They are
determined entirely by the homogeneous quadratic sector \(C_s\), through
\(\mathsf B\). 


\section{Cartan Forms, Curvature, and Heston Characteristic Dynamics}
\label{sec:heston-cartan-characteristics}

We distinguish the left-invariant Cartan form \(\Theta_s\) of the finite
symplectic sector from the full Heston Poincar\'e--Cartan form \(\Theta\).
The holonomy contribution associated with \(C_H\) converts the former into
the latter, whose curvature determines the Heston characteristic field.
Detailed derivations are given in Appendix~\ref{app:invariant-cartan-geometry}.

\subsection{Cartan Form and Curvature of the Finite Symplectic Sector}
\label{sec:heston-finite-cartan}

The Cartan form \(\Theta_s\)  extracts the central component of a
left-invariant vector field. \(\Theta_s\)
is normalized on the central direction and
horizontal on the noncentral left-invariant generators:
\begin{equation*}
\Theta_s(\Xi)=1,\qquad
\Theta_s(\boldsymbol L_{\mathbf a}^{\,s})=0,\qquad
\Theta_s(L_t^s)=0.
\end{equation*}

Using \eqref{eq:heston-Cs-symplectic-form},
\begin{equation}
\begin{aligned}
\Theta_s
& =
\frac{d\zeta}{\zeta}
-\frac12\mathbf a^TJ_4\,d\mathbf a
+C_s\,dt =  \\
& \qquad \qquad 
\frac{d\zeta}{\zeta}
+\frac12\mathbf p^Td\mathbf x
-\frac12\mathbf x^Td\mathbf p
+C_s\,dt.
\label{eq:heston-finite-cartan-form}
\end{aligned}
\end{equation}

The corresponding curvature is

\begin{equation}
\begin{aligned}
\omega_s
& =
d\Theta_s
=
-\frac12\,d\mathbf a^T J_4\wedge d\mathbf a
+dC_s\wedge dt  = \\
& \qquad \qquad 
-d\mathbf x^T\wedge d\mathbf p
+dC_s\wedge dt. 
\label{eq:heston-finite-curvature}
\end{aligned}
\end{equation}

\subsection{Holonomy Contribution and Heston Poincar\'e--Cartan Form}
\label{sec:heston-full-pc-form}

The sector \(C_H\) acts through the multiplicative holonomy
\eqref{eq:heston-affine-holonomy}. For a fixed-time spatial path, \(dt=0\),
and therefore the holonomy is trivial. Thus it does not alter the spatial
left-invariant generators of the finite symplectic Lie-group sector.

For an infinitesimal time displacement,
\begin{equation*}
\mathcal H=1-C_H\,dt+o(dt).
\end{equation*}
The corresponding infinitesimal action on the positive central fiber is
therefore \(-C_H\Xi\). Adding this vertical term to the finite-sector time
generator gives the affine time generator
\begin{equation}
L_t^A
=
L_t^s-C_H \, \Xi
=
\partial_t+(K_s\mathbf a)^T\nabla_{\mathbf a}-C_H \, \Xi.
\label{eq:heston-affine-time-lift}
\end{equation}

The Heston Poincar\'e--Cartan form \(\Theta\) is normalized on the
central generator and horizontal on the spatial left-invariant generators
and on \(L_t^A\). In phase-space variables
\begin{equation}
\begin{aligned}
\Theta
& =
\Theta_s+C_H\,dt  = \\
& \qquad \qquad 
\frac{d\zeta}{\zeta}
+\frac12\mathbf p^T d\mathbf x
-\frac12\mathbf x^T d\mathbf p
+C_A\,dt.
\label{eq:heston-poincare-cartan-form}
\end{aligned}
\end{equation}

The Heston Poincar\'e--Cartan curvature is
\begin{equation}
\omega
=
d\Theta
=
-d\mathbf x^T\wedge d\mathbf p
+dC_A\wedge dt.
\label{eq:heston-full-curvature}
\end{equation}

\subsection{Holonomy-Induced Enlargement of the Generator Algebra}
\label{sec:heston-holonomy-commutators}

The affine time generator \eqref{eq:heston-affine-time-lift} also changes
the commutator structure of the finite-sector generators. In particular,
\begin{equation}
[L_t^A,\boldsymbol L_{\mathbf a}^s]
=
[L_t^s,\boldsymbol L_{\mathbf a}^s]
+
(\nabla_{\mathbf a}C_H)\,\Xi.
\label{eq:affine-enlarged-commutator}
\end{equation}

Thus the finite symplectic Lie algebra itself is unchanged, while
adjoining \(L_t^A\) enlarges the differential-operator algebra by central
terms whose coefficients are derivatives of \(C_H\).

For the Heston complementary sector,
\begin{equation}
\nabla_{\mathbf x}C_H
=
\begin{pmatrix}
0\\
\frac12\mathbf p^T\mathsf A_v\mathbf p
\end{pmatrix},
\qquad
\nabla_{\mathbf p}C_H
=
\mathbf b+v\mathsf A_v\mathbf p.
\label{eq:CH-spatial-gradients}
\end{equation}
Accordingly,
\begin{equation}
[L_t^A,\boldsymbol L_{\mathbf x}^s]
=
[L_t^s,\boldsymbol L_{\mathbf x}^s]
+
(\nabla_{\mathbf x}C_H)\,\Xi,
\qquad
[L_t^A,\boldsymbol L_{\mathbf p}^s]
=
[L_t^s,\boldsymbol L_{\mathbf p}^s]
+
(\nabla_{\mathbf p}C_H)\,\Xi.
\label{eq:affine-enlarged-commutators-blocks}
\end{equation}

The finite symplectic generators close with constant structure
coefficients, whereas adjoining \(L_t^A\) introduces phase-space-dependent
structure functions. Explicit dependence on the variance coordinate occurs
only in the commutator with the momentum generators, through
\(\mathbf b+v\mathsf A_v\mathbf p\). This passage from structure constants
to structure functions reflects the groupoid description: the finite
symplectic Lie group encodes the \(C_s\) transport, while the groupoid
organizes path composition and the multiplicative holonomy associated with
\(C_H\).

The limit \(v\to0\) is regular: although the stochastic covariance
\(v\mathsf A_v\) degenerates at \(v=0\), the affine generator and holonomy
structure remain nonsingular.


\subsection{Poincar\'e--Cartan Characteristic Module}
\label{sec:heston-characteristic-field}

The characteristic module \(\mathcal C_\Theta\) consists of the vector fields that are horizontal
for \(\Theta\) and lie in the kernel of its curvature.
\begin{equation}
\mathcal C_\Theta
=
\left\{
X:\Theta(X)=0,\qquad
\omega(X,\cdot)=0
\right\}.
\label{eq:heston-characteristic-module}
\end{equation}

The characteristic fields \(X_\Theta\in\mathcal C_\Theta\) are geometric
objects associated with the Poincar\'e--Cartan form. Their integral curves
live in the extended phase space and should not be interpreted as sample paths
of the Heston diffusion.

A detailed calculation of \(X_\Theta\) is given in
Appendix~\ref{app:characteristic-module}. For the Heston Poincar\'e--Cartan
form \eqref{eq:heston-poincare-cartan-form}, the defining conditions
\eqref{eq:heston-characteristic-module} leave one independent direction, so
\(\mathcal C_\Theta\) has rank one. Normalizing its generator by
\(dt(X_\Theta)=1\), we obtain, in phase-space coordinates,

\begin{equation}
X_\Theta
=
\partial_t
+\mathbf R^T\nabla_{\mathbf p}
-\bigl(\nabla_{\mathbf p}C_A\bigr)^T\nabla_{\mathbf x}
+\eta\,\Xi
\label{eq:heston-characteristic-field-full}
\end{equation}
where
\begin{equation}
\eta
=
\frac12\left(
\mathbf p^T\nabla_{\mathbf p}
+\mathbf x^T\nabla_{\mathbf x}
-2
\right)C_A.
\label{eq:heston-eta-compact}
\end{equation}

In terms of the left-invariant fields of the finite symplectic Lie-group sector, the
characteristic field \eqref{eq:heston-characteristic-field-full} can be written as
\begin{equation}
\begin{aligned}
X_\Theta
&=
L_t^s
-\bigl(\nabla_{\mathbf p}C_H\bigr)^T
\boldsymbol L_{\mathbf x}^{\,s}
+\bigl(\nabla_{\mathbf x}C_H\bigr)^T
\boldsymbol L_{\mathbf p}^{\,s}
-C_H\Xi.
\end{aligned}
\label{eq:heston-characteristic-left-fields-phase}
\end{equation}

In this representation, the contribution of the homogeneous quadratic sector
\(C_s\) is already contained in the finite-sector time generator \(L_t^s\);
only the complementary contribution \(C_H\) appears explicitly.



\section{Spatial Polarizations of the Quadratic Sector}
\label{sec:heston-spatial-polarization}

A spatial polarization selects the representation variables by eliminating
half of the phase-space variables. 

In standard GAQ, a complete first-order polarization is a maximal
horizontal left subalgebra containing the characteristic module \(\mathcal C_\Theta\). In the
present AHGQ construction, spatial reduction is treated separately from
the characteristic dynamics. We therefore consider subalgebras that are
maximal among the spatial generators of the finite symplectic sector,
horizontal with respect to \(\Theta_s\), and closed under commutators.

The state generators commute among themselves, as do the momentum
generators. They define the two complementary spatial polarization
subalgebras
\begin{equation}
\mathcal P_{\mathrm{mom}}^s
=
\operatorname{span}_{\mathbb R}\{L_x^s,L_v^s\},
\qquad
\mathcal P_{\mathrm{coord}}^s
=
\operatorname{span}_{\mathbb R}\{L_{p_x}^s,L_{p_v}^s\}.
\label{eq:heston-spatial-polarizations}
\end{equation}

These spatial subalgebras are maximal among the spatial generators but are not
complete first-order polarizations of the finite symplectic Lie-group
sector. Since the commutators of \(L_t^s\) with each spatial subalgebra
close within that subalgebra, \(L_t^s\) could be adjoined to obtain the
corresponding GAQ first-order polarization of the quadratic sector. We do not
do so here, because \(L_t^s\) generates only the \(C_s\) evolution. The
full Heston characteristic direction, including the contribution of
\(C_H\), is treated separately through \(X_\Theta\).

\subsection{Polarized Functions}
\label{sec:heston-polarized-functions}

The polarized spaces are constructed from the space of smooth
complex-valued functions on the principal bundle
\(\widetilde G^s\to G^s\),
\begin{equation}
\mathcal F=C^\infty(\widetilde G^s,\mathbb C).
\label{eq:heston-function-space}
\end{equation}

For a spatial polarization subalgebra \(\mathcal P\), define
\begin{equation}
\mathcal F_{\mathcal P}
=
\left\{
\Psi\in\mathcal F:
X\Psi=0\ \text{for every}\ X\in\mathcal P,
\qquad
\Xi\Psi=\Psi
\right\}.
\label{eq:heston-polarized-function-space}
\end{equation}

The condition \(\Xi\Psi=\Psi\) imposes degree-one homogeneity in the
positive central coordinate. Different choices of \(\mathcal P\) give
different representations of the same finite symplectic Lie-group sector.



\subsection{Momentum Polarization}
\label{sec:heston-momentum-polarization}

For a momentum-polarized function
\(\Psi_{\mathrm{mom}}\in\mathcal F_{\mathcal P_{\mathrm{mom}}^s}\),
the polarization conditions are

\begin{equation*}
L_x^s\Psi_{\mathrm{mom}}=0,
\qquad
L_v^s\Psi_{\mathrm{mom}}=0,
\qquad
\Xi\Psi_{\mathrm{mom}}=\Psi_{\mathrm{mom}}.
\end{equation*}

The last condition makes \(\Psi_{\mathrm{mom}}\) linear in the positive
central coordinate \(\zeta\), while the first two determine its dependence
on the state variables \(x\) and \(v\). Their general solution is

\begin{equation}
\Psi_{\mathrm{mom}}
=
\zeta\exp\!\left(\frac12\mathbf x^T\mathbf p\right)
\chi(t,p_x,p_v),
\label{eq:heston-momentum-polarized-function}
\end{equation}

where \(\chi(t,p_x,p_v)\) contains the remaining dependence and defines
the reduced momentum-space function.

The characteristic evolution preserves the momentum-polarized space. We have

\begin{equation}
[X_\Theta,L_{x_j}^s]
=
\sum_i\frac{\partial R_j}{\partial p_i}L_{x_i}^s
\in\mathcal P_{\mathrm{mom}}^s,
\qquad
[X_\Theta,\Xi]=0.
\label{eq:heston-characteristic-polarization-commutator}
\end{equation}

Therefore, if
\(\Psi_{\mathrm{mom}}\in\mathcal F_{\mathcal P_{\mathrm{mom}}^s}\),
then

\begin{equation*}
\begin{aligned}
L_{x_j}^s(X_\Theta\Psi_{\mathrm{mom}})
&=
X_\Theta(L_{x_j}^s\Psi_{\mathrm{mom}})
-
[X_\Theta,L_{x_j}^s]\Psi_{\mathrm{mom}}
=0,
\\
\Xi(X_\Theta\Psi_{\mathrm{mom}})
&=
X_\Theta(\Xi\Psi_{\mathrm{mom}})
-
[X_\Theta,\Xi]\Psi_{\mathrm{mom}}
=
X_\Theta\Psi_{\mathrm{mom}},
\end{aligned}
\end{equation*}

which implies that $X_\Theta\Psi_{\mathrm{mom}} \in \mathcal F_{\mathcal P_{\mathrm{mom}}^s}$.

\subsection{Momentum Characteristic Operator}
\label{sec:heston-momentum-characteristic-operator}

For a momentum-polarized function, the dependence on \(\zeta\), \(x\),
and \(v\) is fixed by \eqref{eq:heston-momentum-polarized-function}.
We therefore define the momentum reduction map by
\begin{equation}
\mathcal R_{\mathrm{mom}}\Psi
=
\zeta^{-1}
\exp\!\left(-\frac12\mathbf x^T\mathbf p\right)\Psi,
\qquad
\mathcal R_{\mathrm{mom}}\Psi_{\mathrm{mom}}=\chi,
\qquad
\chi=\chi(t,p_x,p_v).
\label{eq:heston-momentum-reduction-map}
\end{equation}

Since \(X_\Theta\) preserves the momentum-polarized space, its reduced
action is defined by
\begin{equation*}
X_\Theta^{\mathrm{mom}}\chi
=
\mathcal R_{\mathrm{mom}}
\left(X_\Theta\Psi_{\mathrm{mom}}\right).
\end{equation*}

Recall that the nonzero component of the vector-valued Riccati symbol is
\begin{equation}
R
=
\frac12(p_x^2-p_x)
+(\rho\sigma_\nu p_x-\kappa)p_v
+\frac12\sigma_\nu^2p_v^2.
\label{eq:heston-R-momentum}
\end{equation}

Substituting
\eqref{eq:heston-momentum-polarized-function} into
\eqref{eq:heston-characteristic-field-full}, with
\(C_A=F+\mathbf x^T\mathbf R\), gives
\begin{equation*}
\mathcal R_{\mathrm{mom}}
\left(X_\Theta\Psi_{\mathrm{mom}}\right)
=
\left[
\partial_t
+\mathbf R^T\nabla_{\mathbf p}
+\frac12\mathbf x^T\mathbf R
-\frac12\mathbf p^T\nabla_{\mathbf p}C_A
+\eta
\right]\chi.
\end{equation*}
Using
\begin{equation*}
\eta
=
\frac12
\left(
\mathbf p^T\nabla_{\mathbf p}
+\mathbf x^T\nabla_{\mathbf x}
-2
\right)C_A,
\qquad
\nabla_{\mathbf x}C_A=\mathbf R,
\end{equation*}
the state-dependent terms cancel, leaving
\begin{equation}
X_\Theta^{\mathrm{mom}}
=
\partial_t+R\,\partial_{p_v}-F.
\label{eq:heston-reduced-momentum-characteristic}
\end{equation}

Since
\(F=(r-\delta)p_x+\kappa\theta p_v-r\), this becomes
\begin{equation}
X_\Theta^{\mathrm{mom}}
=
\partial_t
+R\,\partial_{p_v}
-(r-\delta)p_x-\kappa\theta p_v+r.
\label{eq:heston-reduced-momentum-characteristic-expanded}
\end{equation}

Thus momentum polarization removes the state variables from the
characteristic operator, leaving the Riccati component \(R\) and the
state-independent affine term \(F\).


\subsection{Coordinate Polarization and a Coordinate Pricing Operator}
\label{sec:heston-coordinate-representation}

The coordinate-polarized function space
\(\mathcal F_{\mathrm{coord}}:=\mathcal F_{\mathcal P_{\mathrm{coord}}^s}\)
is defined by
\begin{equation}
\mathcal F_{\mathrm{coord}}
=
\left\{
\Psi\in\mathcal F:
L_{p_x}^s\Psi=0,\quad
L_{p_v}^s\Psi=0,\quad
\Xi\Psi=\Psi
\right\}.
\label{eq:heston-coordinate-polarization}
\end{equation}
The polarization conditions give
\begin{equation}
\Psi_{\mathrm{coord}}
=
\zeta\exp\!\left(-\frac12\mathbf x^T\mathbf p\right)V(t,\mathbf x),
\qquad
\Psi_{\mathrm{coord}}\in\mathcal F_{\mathrm{coord}}.
\label{eq:heston-coordinate-polarized-function}
\end{equation}
Thus the dependence on \(\zeta\) and \(\mathbf p\) is fixed by the
polarization, while \(V(t,\mathbf x)\) is the reduced coordinate
representation.

Removing the symplectic transport from the spatial right-invariant
generators gives
\begin{equation}
M_s(t)^{-T}\boldsymbol R_{\mathbf a}^{\,s}
=
\begin{pmatrix}
\mathbf P\\
-\mathbf X
\end{pmatrix},
\qquad
\mathbf P
=
\nabla_{\mathbf x}+\frac12\mathbf p\,\Xi,
\qquad
\mathbf X
=
-\nabla_{\mathbf p}+\frac12\mathbf x\,\Xi.
\label{eq:heston-right-generators-PX}
\end{equation}

The operators \(\mathbf P\) and \(\mathbf X\) preserve the
coordinate-polarized space because they are linear combinations of
right-invariant generators. Their action on \(\Psi_{\mathrm{coord}}\) is
\begin{equation}
\mathbf P\Psi_{\mathrm{coord}}
=
\zeta e^{-\frac12\mathbf x^T\mathbf p}\nabla_{\mathbf x}V,
\qquad
\mathbf X\Psi_{\mathrm{coord}}
=
\zeta e^{-\frac12\mathbf x^T\mathbf p}\mathbf x\,V.
\label{eq:heston-PX-coordinate-action}
\end{equation}
Hence, on the reduced coordinate representation,
\begin{equation}
\mathbf P\longmapsto\nabla_{\mathbf x},
\qquad
\mathbf X\longmapsto\mathbf x,
\qquad
\Xi\longmapsto1,
\qquad
[\mathbf P,\mathbf X^T]\longmapsto I_2.
\label{eq:heston-PX-coordinate-reduction}
\end{equation}

The polarization conditions are imposed with left-invariant generators,
whereas right-invariant generators act on the resulting polarized
functions. We therefore use the right action to represent the affine Heston symbol
\(C_A\) on the coordinate-polarized space. Its reduction will be shown to
coincide with the standard Heston pricing operator.

The operators \(\mathbf P\) and \(\mathbf X\) represent, under this right
action, the phase-space variables \(\mathbf p\) and \(\mathbf x\).
Accordingly, from
\begin{equation*}
C_A(\mathbf x,\mathbf p)
=
F(\mathbf p)+\mathbf x^T\mathbf R(\mathbf p),
\end{equation*}
we define the corresponding right-operator representative by
\begin{equation}
\widehat C_A^{\,R}
=
F(\mathbf P)+\mathbf X^T\mathbf R(\mathbf P).
\label{eq:heston-affine-right-operator}
\end{equation}
Its reduction on coordinate-polarized functions gives the Heston
coordinate-space pricing operator. No additional ordering prescription is
required, since the ordering is inherited directly from the affine
decomposition of \(C_A\).

Using \eqref{eq:heston-PX-coordinate-action},
\begin{equation}
\widehat C_A^{\,R}\Psi_{\mathrm{coord}}
=
\zeta e^{-\frac12\mathbf x^T\mathbf p}
\left[
F(\nabla_{\mathbf x})
+\mathbf x^T\mathbf R(\nabla_{\mathbf x})
\right]V
=
\zeta e^{-\frac12\mathbf x^T\mathbf p}\mathcal L_A V.
\label{eq:heston-affine-right-operator-action}
\end{equation}
The resulting reduced coordinate operator is then
\begin{equation}
\mathcal L_A
=
F(\nabla_{\mathbf x})
+\mathbf x^T\mathbf R(\nabla_{\mathbf x})
\label{eq:heston-coordinate-operator}
\end{equation}

In Heston parameters,
\begin{equation}
\mathcal L_A
=
\left(r-\delta-\frac12v\right)\partial_x
+\kappa(\theta-v)\partial_v
+\frac12v\left(
\partial_x^2
+2\rho\sigma_\nu\partial_x\partial_v
+\sigma_\nu^2\partial_v^2
\right)
-r.
\label{eq:heston-coordinate-generator}
\end{equation}

Preservation of the coordinate polarization does not determine the pricing
operator uniquely. Any differential operator generated by the
right-invariant fields preserves \(\mathcal F_{\mathrm{coord}}\). The
representative \eqref{eq:heston-affine-right-operator} is distinguished here
because it applies the canonical operators
\eqref{eq:heston-right-generators-PX} directly to the affine symbol \(C_A\).
The associated diffusion generator and Heston stochastic equations are recovered
in Appendix~\ref{app:heston-diffusion-geometry}.


\section{Financial Applications: Mellin Pricing, Projective Propagation, and Numerical Validation}
\label{sec:financial-applications}

The preceding sections construct the momentum representation directly from
the Heston Poincar\'e--Cartan characteristic field. We now use that
representation to price European instruments. Mellin transformation supplies
the terminal momentum-space data, while the reduced characteristic dynamics
propagates the corresponding momentum components. This construction is
equivalent to the usual Laplace--Fourier affine representation; see, for
example, \cite{CarrMadan,DuffiePanSingleton2000,Mellin0,Mellin2,Mellin4}.
\subsection{Propagation of Momentum-Polarized Functions and Momentum Propagator}
\label{subsec:momentum-pricing-kernel}

Momentum-polarized functions provide the momentum-space representation
of European financial instruments. At maturity, the momentum-polarized
function is determined by the transform of the payoff. Its evolution
away from maturity is governed by the reduced momentum characteristic
operator, while the price in the state variables is recovered by an
inverse integral transform.

The Mellin representation is particularly natural in the present
construction. The central extension by \(\mathbb R_+\), rather than
\(U(1)\), leads naturally to real exponential momentum modes \(e^{p_xx}\).
Since \(x=\log S\), these modes become powers \(S^{p_x}\), and the
bilateral Laplace representation in \(x\) is equivalently a Mellin
representation in the positive stock variable \(S\).

We proceed in three steps. We first determine the characteristic flow
of the reduced momentum operator. We then introduce a momentum
propagator and apply the reduced operator to the propagated
momentum-polarized function. Finally, compatibility with the
characteristic flow determines the propagator and allows the price to
be recovered by inverse Mellin transformation.

Recall that the reduced momentum characteristic operator is

\begin{equation}
X_\Theta^{\mathrm{mom}}
=
\partial_\tau+R(p_x,p_v)\partial_{p_v}-F(p_x,p_v).
\label{eq:heston-momentum-characteristic-pricing}
\end{equation}

Here

\begin{equation*}
R(p_x,p_v)
=
\frac12(p_x^2-p_x)
+
(\rho\sigma_\nu p_x-\kappa)p_v
+
\frac12\sigma_\nu^2p_v^2,
\qquad
F(p_x,p_v)
=
(r-\delta)p_x+\kappa\theta p_v-r.
\end{equation*}

The reduced momentum-polarized function satisfies

\begin{equation}
X_\Theta^{\mathrm{mom}}\chi=0.
\label{eq:heston-reduced-momentum-equation-pricing}
\end{equation}

The characteristic equations associated with this first-order equation
are

\begin{equation}
\frac{dp_x}{d\tau}=0,
\qquad
\frac{dp_v}{d\tau}=R(p_x,p_v),
\qquad
\frac{d\chi}{d\tau}=F(p_x,p_v)\chi.
\label{eq:heston-financial-momentum-flow}
\end{equation}

The first equation shows that the log-price momentum is conserved.
For a Mellin component we write

\begin{equation*}
p_x=-q.
\end{equation*}

Initially \(q\in\mathbb R\); it is later complexified along the Mellin
contour.

We consider European payoffs depending only on the terminal stock price.
Consequently, their terminal momentum representation has no variance
momentum, and

\begin{equation*}
p_v(0)=0.
\end{equation*}

Writing

\begin{equation*}
p_v(\tau)=D(\tau,q),
\qquad
D(0,q)=0,
\end{equation*}

the second characteristic equation in
\eqref{eq:heston-financial-momentum-flow} becomes

\begin{equation}
\frac{dD}{d\tau}
=
\frac12\sigma_\nu^2D^2
-
(\rho\sigma_\nu q+\kappa)D
+
\frac12(q^2+q),
\qquad
D(0,q)=0.
\label{eq:heston-riccati-mellin}
\end{equation}

Thus the Heston Riccati equation is the characteristic equation for the
variance momentum. Although the terminal variance momentum vanishes,
the characteristic flow generates a nonzero
\(p_v=D(\tau,q)\) for \(\tau>0\).

We next introduce the momentum propagator. Let
\(\widehat\Phi(q)\) denote the terminal value of the reduced momentum-space
function, identified with the Mellin transform of the European payoff. Along
the characteristic labelled by \(q\), define

\begin{equation*}
\chi_q(\tau)
:=
\chi\bigl(\tau,-q,D(\tau,q)\bigr).
\end{equation*}

We write the propagated momentum-polarized function as

\begin{equation}
\chi_q(\tau)
=
\int_\Gamma
G(\tau;q,q')\,\widehat\Phi(q')\,dq',
\label{eq:heston-momentum-propagator-representation}
\end{equation}

where \(G(\tau;q,q')\) is the momentum propagator. At maturity no
propagation has taken place, so the identity condition is

\begin{equation}
G(0;q,q')=\delta(q-q').
\label{eq:heston-momentum-propagator-initial}
\end{equation}

We now apply the reduced momentum equation. Along the characteristic
\(p_x=-q\), \(p_v=D(\tau,q)\), the flow of the reduced function is

\begin{equation}
\frac{d\chi_q}{d\tau}
=
F\bigl(-q,D(\tau,q)\bigr)\chi_q.
\label{eq:heston-momentum-equation-along-characteristic}
\end{equation}

Substitution of
\eqref{eq:heston-momentum-propagator-representation} gives

\begin{equation*}
\int_\Gamma
\widehat\Phi(q')
\left[
\partial_\tau G(\tau;q,q')
-
F\bigl(-q,D(\tau,q)\bigr)G(\tau;q,q')
\right]dq'
=
0.
\end{equation*}

Since the terminal momentum data \(\widehat\Phi(q')\) are arbitrary,
the propagator must satisfy

\begin{equation}
\partial_\tau G(\tau;q,q')
=
F\bigl(-q,D(\tau,q)\bigr)G(\tau;q,q'),
\qquad
G(0;q,q')=\delta(q-q').
\label{eq:heston-momentum-propagator-equation}
\end{equation}

Since \(p_x=-q\) is conserved, distinct Mellin momenta do not mix under
the evolution. The propagator is therefore diagonal in \(q\),

\begin{equation}
G(\tau;q,q')
=
\delta(q-q')e^{A(\tau,q)}.
\label{eq:heston-momentum-propagator}
\end{equation}

Substitution into
\eqref{eq:heston-momentum-propagator-equation} gives

\begin{equation}
\frac{dA}{d\tau}
=
F\bigl(-q,D(\tau,q)\bigr)
=
-(r-\delta)q+\kappa\theta D(\tau,q)-r,
\qquad
A(0,q)=0.
\label{eq:heston-affine-amplitude-mellin}
\end{equation}

Equivalently,

\begin{equation}
A(\tau,q)
=
\int_0^\tau
F\bigl(-q,D(s,q)\bigr)\,ds.
\label{eq:heston-affine-amplitude-integral}
\end{equation}

The characteristic flow transports the variance momentum from
\(p_v=0\) to \(p_v=D(\tau,q)\), while the momentum propagator multiplies
the corresponding Mellin component by \(e^{A(\tau,q)}\).

\subsubsection{Instrument Price in the State Variables}

The transported momentum is

\begin{equation*}
\mathbf p(\tau,q)=\bigl(-q,D(\tau,q)\bigr)^T.
\end{equation*}

Hence the corresponding exponential factor in the state variables is

\begin{equation*}
e^{\mathbf x^T\mathbf p}
=
e^{-qx+D(\tau,q)v}
=
S^{-q}e^{D(\tau,q)v}.
\end{equation*}

Let \(t\) denote calendar valuation time, \(T\) the maturity date, and
\(\tau=T-t\). The instrument price is obtained by inverse Mellin
transformation of the propagated momentum-space representation,

\begin{equation}
C(\tau,S,v)
=
\frac{1}{2\pi\iu}
\int_\Gamma\!\!\int_\Gamma
G(\tau;q,q')\,
\widehat\Phi(q')\,
S^{-q}e^{D(\tau,q)v}\,
dq'\,dq.
\label{eq:heston-mellin-price-propagator-double}
\end{equation}

Using the diagonal form
\eqref{eq:heston-momentum-propagator}, the \(q'\)-integration gives

\begin{equation}
C(\tau,S,v)
=
\frac{1}{2\pi\iu}
\int_\Gamma
\widehat\Phi(q)\,
e^{A(\tau,q)+D(\tau,q)v}
S^{-q}\,dq.
\label{eq:heston-mellin-price}
\end{equation}

At maturity, \(D(0,q)=A(0,q)=0\), so

\begin{equation*}
C(0,S,v)
=
\frac{1}{2\pi\iu}
\int_\Gamma
\widehat\Phi(q)S^{-q}\,dq
=
\Phi(S).
\end{equation*}

The terminal value is therefore independent of \(v\), as required.
The variance dependence of the pre-maturity price is generated entirely
by the characteristic transport \(p_v(0)=0\) to
\(p_v(\tau)=D(\tau,q)\).

The contour \(\Gamma\) is chosen inside the intersection of the Mellin
convergence strip of the terminal payoff and the domain in which the
Heston Riccati flow remains finite for the maturity under consideration
\cite{KellerRessel2011,KellerResselMayerhofer2015}.


\FloatBarrier


\subsubsection{Projective Representation of the Riccati Flow}
\label{subsec:projective-riccati}

The nonlinear Riccati characteristic obtained above admits a
two-dimensional projective linearization. Writing

\begin{equation*}
\frac{dD}{d\tau}=\alpha D^2+\beta D+\gamma,
\qquad
\alpha=\frac12\sigma_\nu^2,
\qquad
\beta=-(\rho\sigma_\nu q+\kappa),
\qquad
\gamma=\frac12(q^2+q),
\end{equation*}

and introducing the projective coordinate \(D=u/w\), the Riccati
equation is generated by the linear system

\begin{equation}
\frac{d}{d\tau}
\begin{pmatrix}u\\w\end{pmatrix}
=
A_R(q)
\begin{pmatrix}u\\w\end{pmatrix},
\qquad
A_R(q)
=
\begin{pmatrix}
\frac12\beta&\gamma\\
-\alpha&-\frac12\beta
\end{pmatrix}.
\label{eq:projective-AR}
\end{equation}

Since \(A_R(q)\) has zero trace, for real \(q\) it belongs to
\(\mathfrak{sl}(2,\mathbb R)\). Along a complex Mellin contour, the
same construction is understood in the complexified algebra. The
corresponding linear transport is

\begin{equation*}
M_R(\tau,q)=e^{\tau A_R(q)}=(m_{ij}(\tau,q)).
\end{equation*}

For real \(q\), \(M_R(\tau,q)\in SL(2,\mathbb R)\). The initial
condition \(D(0,q)=0\) gives

\begin{equation}
D(\tau,q)=\frac{m_{12}(\tau,q)}{m_{22}(\tau,q)}.
\label{eq:D-projective-ratio}
\end{equation}

Thus the nonlinear Heston Riccati characteristic is the projective image of
a two-dimensional linear flow. This auxiliary projective transport is
distinct from the four-dimensional symplectic transport \(M_s(t)\):
\(M_s(t)\) belongs to the finite symplectic Lie-group sector, whereas
\(M_R(\tau,q)\) arises after momentum polarization and Mellin reduction.

Formulas and derivations for the Riccati characteristic \(D\) and the
affine amplitude \(A\) are given in
Appendix~\ref{app:projective-riccati}. For later reference, we record the
closed expressions obtained from this projective representation. Define

\begin{equation*}
d_q
=
\sqrt{(\kappa+\rho\sigma_\nu q)^2-\sigma_\nu^2(q^2+q)},
\qquad
E_\tau(q)
=
\cosh\!\left(\frac{d_q\tau}{2}\right)
+
\frac{\kappa+\rho\sigma_\nu q}{d_q}
\sinh\!\left(\frac{d_q\tau}{2}\right).
\end{equation*}

The variance-momentum characteristic \(D(\tau,q)=p_v(\tau,q)\) is

\begin{equation}
D(\tau,q)
=
\frac{(q^2+q)\sinh\!\left(\frac{d_q\tau}{2}\right)}
{d_q\cosh\!\left(\frac{d_q\tau}{2}\right)
+(\kappa+\rho\sigma_\nu q)\sinh\!\left(\frac{d_q\tau}{2}\right)},
\qquad
D(0,q)=0.
\label{eq:heston-D-final}
\end{equation}

For \(\sigma_\nu\neq0\), the corresponding affine amplitude is

\begin{equation}
A(\tau,q)
=
-\bigl[(r-\delta)q+r\bigr]\tau
+
\frac{\kappa\theta(\kappa+\rho\sigma_\nu q)}{\sigma_\nu^2}\tau
-
\frac{2\kappa\theta}{\sigma_\nu^2}\log E_\tau(q),
\qquad
A(0,q)=0.
\label{eq:heston-A-final}
\end{equation}

Along a complex Mellin contour,
the square root defining \(d_q\) and, in particular, the logarithm
\(\log E_\tau(q)\) must be evaluated consistently. The continuous choice of the logarithm along the contour enters the
numerical validation below.

\subsection{Numerical Validation}
\label{sec:heston_logform_validation}

The Mellin--projective construction above expresses European prices in
terms of the Riccati characteristic \(D\) and the affine amplitude \(A\).
We validate this representation numerically by comparing European call
prices with the standard semi-analytic Heston formula
\cite{Heston1993,DuffiePanSingleton2000,DuffieFilipovicSchachermayer2003}.
The purpose is a consistency check of the momentum-space and projective
construction, not a rederivation of the standard characteristic-function
formula.

We first record the call-price Mellin integral used in the implementation and
then report the numerical comparison.

\subsubsection{European Call Option Price}
\label{sec:heston-european-prices}

For a European call payoff
\(\Phi_{\mathrm{call}}(S)=(S-K)^+\), the Mellin transform is

\begin{equation}
\widehat\Phi_{\mathrm{call}}(q)
=
\frac{K^{q+1}}{q(q+1)},
\qquad
\Re(q)<-1.
\label{eq:heston-call-mellin-transform}
\end{equation}

For convenience, define the propagated Mellin mode

\begin{equation}
\mathcal K_q(\tau,x,v)
=
\exp\!\left[-qx+D(\tau,q)v+A(\tau,q)\right],
\qquad
\mathcal K_q(0,x,v)=e^{-qx}.
\label{eq:heston-propagated-mellin-mode}
\end{equation}

Using the momentum evolution derived above, the call price is then

\begin{equation}
C_{\mathrm{call}}(t,S,v;K,T)
=
\frac{1}{2\pi\iu}
\int_\Gamma
\frac{K^{q+1}}{q(q+1)}
\mathcal K_q(T-t,\log S,v)\,dq.
\label{eq:mellin-heston-call}
\end{equation}

The contour is parametrized by \(q=c+\iu\eta\), with
\(\eta\in\mathbb R\). The constant \(c\) is chosen with \(c<-1\) and
within the domain in which the Heston Riccati solution remains finite
for the maturity \(T-t\).

\subsubsection{Numerical Tests}
\label{sec:numerical_test}

For the numerical tests we set the valuation time to \(t=0\), so that
\(\tau=T\). We compute the Riccati characteristic from
\eqref{eq:heston-D-final} and the affine amplitude from
\eqref{eq:heston-A-final}. Since \(E_T(q)\) is complex along the inversion
contour, the implementation uses a continuous branch of
\(\log E_T(q)\), anchored at the center of the contour.

The baseline configuration, with zero dividend yield, is

\begin{equation*}
\begin{aligned}
S_0& =K=100,\quad r=0.03,\quad \delta=0,\quad T=1 \\
\kappa& =2,\quad \theta=0.04,\quad \sigma_\nu=0.35,
\quad \rho=-0.7,\quad v_0=0.04.
\end{aligned}
\end{equation*}

We vary \(v_0\), \(\theta\), \(\kappa\), \(\sigma_\nu\), \(\rho\), and
\(T\) one parameter at a time, keeping the remaining parameters fixed.
The resulting European call prices are compared with the standard
semi-analytic Heston benchmark in
Table~\ref{tab:heston_centered_log_stress_ranges}.

\begin{table}[htbp]
\centering
\scriptsize
\setlength{\tabcolsep}{4pt}
\renewcommand{\arraystretch}{1.08}
\begin{tabular}{@{}lccccc@{}}
\toprule
Varied parameter
& Values tested
& Benchmark price range
& Mellin--projective price range
& Max. abs. diff.
& Feller range \\
\midrule
\(v_0\) & \(0.01\)--\(0.09\) & \(7.85571767\)--\(10.99728901\) & \(7.85571767\)--\(10.99728903\) & \(1.89{\times}10^{-8}\) & \(0.0375\) \\
\(\theta\) & \(0.01\)--\(0.09\) & \(7.26603514\)--\(11.56109741\) & \(7.26603499\)--\(11.56109743\) & \(1.47{\times}10^{-7}\) & \(-0.0825\)--\(0.2375\) \\
\(\kappa\) & \(0.50\)--\(5.00\) & \(8.87483850\)--\(9.34310020\) & \(8.87483854\)--\(9.34310021\) & \(4.26{\times}10^{-8}\) & \(-0.0825\)--\(0.2775\) \\
\(\sigma_\nu\) & \(0.10\)--\(1.00\) & \(8.01630215\)--\(9.40634453\) & \(8.01631142\)--\(9.40634455\) & \(9.27{\times}10^{-6}\) & \(-0.8400\)--\(0.1500\) \\
\(\rho\) & \(-0.90\)--\(0.30\) & \(9.12143406\)--\(9.18078083\) & \(9.12143407\)--\(9.18078084\) & \(1.30{\times}10^{-8}\) & \(0.0375\) \\
\(T\) & \(0.50\)--\(10.00\) & \(6.22147748\)--\(36.88027124\) & \(6.22147749\)--\(36.88027137\) & \(1.29{\times}10^{-7}\) & \(0.0375\) \\
\bottomrule
\end{tabular}
\caption{Range summary. The benchmark is the standard semi-analytic Heston formula. The price ranges are the minimum and maximum prices observed over the tested values, and the reported error is the maximum absolute difference between the benchmark and the Mellin--projective implementation. The final column reports the range of the Feller diagnostic \(2\kappa\theta-\sigma_\nu^2\).}
\label{tab:heston_centered_log_stress_ranges}
\end{table}

\begin{figure}[ht]
\centering
\IfFileExists{heston_centered_log_prices_by_case.png}{%
\includegraphics[width=0.75\textwidth]{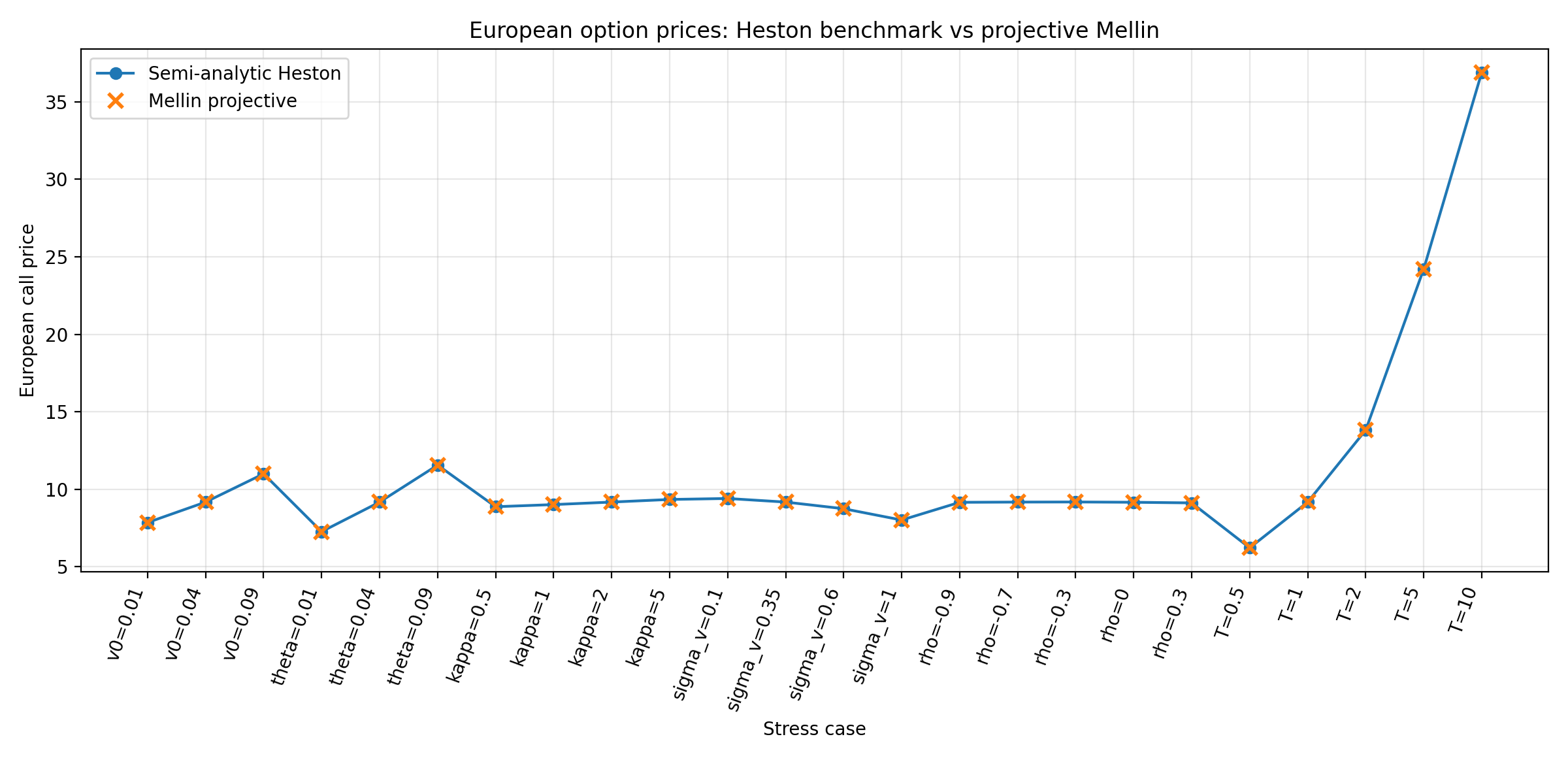}%
}{\fbox{\parbox{0.72\textwidth}{\centering Numerical-validation figure: \\ \texttt{heston\_centered\_log\_prices\_by\_case.png}}}}
\caption{Visual consistency check. The connected curve gives the semi-analytic Heston characteristic-function benchmark, while the cross markers give the Mellin--projective prices.}
\label{fig:heston-numerical-validation}
\end{figure}

\FloatBarrier
\clearpage

\section{Conclusion}
\label{sec:conclusion}

The main contribution of the AHGQ construction is structural: it provides a
common geometric origin for the Heston Riccati dynamics and a coordinate-space
realization of the Heston pricing operator. The affine pricing symbol is separated according to two
geometric roles. The homogeneous quadratic sector \(C_s\) determines the
finite symplectic Lie-group structure, while \(C_H\) is represented through
multiplicative holonomy. Their combination gives the full Heston Poincar\'e--Cartan form and its
characteristic field.

The characteristic field of this structure yields, after spatial
polarization and momentum reduction, the scalar Heston Riccati equation and
the affine amplitude entering the transform representation. Mellin
decomposition then gives the European pricing representation. The resulting
Mellin--projective implementation agrees numerically with the standard
semi-analytic Heston formula, providing a consistency check of the
construction.

The Black--Scholes limit further illustrates that the geometric
decomposition must be recomputed after freezing the variance rather than
obtained by separately restricting the Heston sectors. A natural continuation
is to apply the AHGQ framework to other affine financial models.

\appendix


\appsection{Heston Stochastic Equations}
\label{app:heston-diffusion-geometry}

The drift and covariance data used in the affine construction
recover directly the standard Heston diffusion in log-price and variance
coordinates.

The Heston state-dependent covariance matrix is
\(\mathsf D(v)=v\mathsf A_v\), with \(\mathsf A_v\) given in
\eqref{eq:heston-Av}.

\begin{equation}
\mathsf D(v)
=
v
\begin{pmatrix}
1&\rho\sigma_\nu\\
\rho\sigma_\nu&\sigma_\nu^2
\end{pmatrix}.
\label{eq:heston-Dv}
\end{equation}

A convenient square-root factorization of the covariance matrix is

\begin{equation}
\mathsf D(v)
=
\Sigma(v)\Sigma(v)^T,
\qquad
\Sigma(v)
=
\sqrt v
\begin{pmatrix}
1&0\\
\rho\sigma_\nu&\sigma_\nu\sqrt{1-\rho^2}
\end{pmatrix}.
\label{eq:appendix-heston-covariance-factorization}
\end{equation}

The Heston affine drift is given in  \eqref{eq:heston-affine-drift-data}.

\begin{equation}
\boldsymbol\mu(v)
=
\mathbf b+\mathsf B\mathbf x
=
\begin{pmatrix}
r-\delta-\frac12v\\
\kappa(\theta-v)
\end{pmatrix}.
\label{eq:appendix-heston-drift}
\end{equation}

Let

\begin{equation*}
\mathbf x_t=(x_t,v_t)^T,
\qquad
\mathbf W_t=
\bigl(W_t^{(1)},W_t^{(2)}\bigr)^T,
\end{equation*}

where \(W_t^{(1)}\) and \(W_t^{(2)}\) are independent Brownian motions.
The diffusion associated with
\eqref{eq:heston-Dv} and \eqref{eq:appendix-heston-drift} is therefore

\begin{equation}
d\mathbf x_t
=
\boldsymbol\mu(v_t)\,dt
+
\Sigma(v_t)\,d\mathbf W_t.
\label{eq:appendix-heston-vector-sde}
\end{equation}

In components,

\begin{equation}
\begin{aligned}
dx_t
&=
\left(r-\delta-\frac12v_t\right)dt
+
\sqrt{v_t}\,dW_t^{(1)},
\\
dv_t
&=
\kappa(\theta-v_t)dt
+
\sigma_\nu\sqrt{v_t}
\left(
\rho\,dW_t^{(1)}
+
\sqrt{1-\rho^2}\,dW_t^{(2)}
\right).
\end{aligned}
\label{eq:appendix-heston-component-sde}
\end{equation}

Define the correlated Brownian motions
\(dW_t^{(x)}=dW_t^{(1)}\) and
\(dW_t^{(v)}
=\rho\,dW_t^{(1)}+\sqrt{1-\rho^2}\,dW_t^{(2)}\),
with quadratic covariation
\(dW_t^{(x)}\,dW_t^{(v)}=\rho\,dt\).
The Heston stochastic equations can therefore be written in the
standard correlated form

\begin{equation}
\begin{aligned}
dx_t
&=
\left(r-\delta-\frac12v_t\right)dt
+
\sqrt{v_t}\,dW_t^{(x)},
\\
dv_t
&=
\kappa(\theta-v_t)dt
+
\sigma_\nu\sqrt{v_t}\,dW_t^{(v)}.
\end{aligned}
\label{eq:appendix-heston-standard-sde}
\end{equation}

\appsection{The Black--Scholes Reduction}
\label{app:black-scholes-reduction}

The Black--Scholes limit is obtained by freezing the variance,
\begin{equation}
\sigma_\nu=0,\qquad v=\theta=\bar v,\qquad p_v=0.
\label{eq:bs-reduction-conditions}
\end{equation}
After this reduction, \(v\) is no longer a state coordinate. Therefore the
geometric decomposition must be recomputed from the reduced affine symbol,
rather than obtained by separately restricting \(C_s\) and \(C_H\).
The reduced symbol is
\begin{equation}
C_A^{BS}(p_x)
=
\frac12\bar v p_x^2
+\left(r-\delta-\frac12\bar v\right)p_x-r.
\label{eq:bs-reduced-affine-symbol}
\end{equation}
Its natural quadratic/holonomy decomposition is
\begin{equation}
C_s^{BS}=\frac12\bar v p_x^2,
\qquad
C_H^{BS}=\left(r-\delta-\frac12\bar v\right)p_x-r.
\label{eq:bs-reduced-split}
\end{equation}
Thus the Heston term \(\frac12vp_x^2\) becomes a genuine homogeneous
quadratic term once \(v=\bar v\) is frozen and moves into the
reduced symplectic sector.

With \(\mathbf a_{BS}=(x,p_x)^T\) and
\begin{equation*}
J_2=\begin{pmatrix}0&1\\-1&0\end{pmatrix},
\end{equation*}
we have
\begin{equation}
C_s^{BS}=\frac12\mathbf a_{BS}^TJ_2K_{BS}\mathbf a_{BS},
\qquad
K_{BS}=\begin{pmatrix}0&-\bar v\\0&0\end{pmatrix}.
\label{eq:bs-symplectic-generator}
\end{equation}
Hence
\begin{equation}
M_{BS}(t)=e^{tK_{BS}}
=\begin{pmatrix}1&-\bar v t\\0&1\end{pmatrix}.
\label{eq:bs-symplectic-transport}
\end{equation}
The corresponding symplectic cocycle is
\begin{equation}
\epsilon_2^{BS}(g',g)
=\frac12\bigl(M_{BS}(t)\mathbf a_{BS}'\bigr)^TJ_2\mathbf a_{BS}.
\label{eq:bs-symplectic-cocycle}
\end{equation}
This is the Bargmann-type central extension underlying the
Black--Scholes/Galilean correspondence \cite{bargmann1,bargmann2}. The
remaining drift and discounting contribution is carried separately by
\begin{equation}
\mathcal H_{BS}[\gamma]
=\exp\!\left(-\int_\gamma C_H^{BS}(p_x)\,dt\right).
\label{eq:bs-holonomy}
\end{equation}
Thus the constant-variance limit now emerges directly from the reduced affine
symbol: the diffusion coefficient \(\bar v\) generates the symplectic shear
and nontrivial central cocycle, while only the first-order drift and discount
remain in the holonomy sector.
\appsection{Explicit Projective Solution of the Heston Riccati Flow}
\label{app:projective-riccati}

This appendix gives the explicit solution of the projective Riccati
construction introduced in Section~\ref{subsec:projective-riccati}.
The purpose is to obtain closed expressions for the Riccati
characteristic \(D(\tau,q)\) and for the amplitude $A(\tau,q)$.

After Mellin reduction, the variance-momentum characteristic satisfies

\begin{equation*}
\frac{dD}{d\tau}=\alpha D^2+\beta D+\gamma,
\qquad
\alpha=\frac12\sigma_\nu^2,
\qquad
\beta=-(\rho\sigma_\nu q+\kappa),
\qquad
\gamma=\frac12(q^2+q),
\qquad
D(0,q)=0.
\end{equation*}

As discussed in Section~\ref{subsec:projective-riccati}, introduce
\(D=u/w\) and the two-dimensional linear system

\begin{equation*}
\frac{d}{d\tau}
\begin{pmatrix}u\\w\end{pmatrix}
=
A_R(q)
\begin{pmatrix}u\\w\end{pmatrix},
\qquad
A_R(q)
=
\begin{pmatrix}
\frac12\beta&\gamma\\
-\alpha&-\frac12\beta
\end{pmatrix}.
\end{equation*}

The two equations for \(u\) and \(w\) are therefore

\begin{equation*}
\dot u=\frac12\beta u+\gamma w,
\qquad
\dot w=-\alpha u-\frac12\beta w.
\end{equation*}

Differentiating the projective coordinate \(D=u/w\) gives

\begin{equation}
\frac{dD}{d\tau}
=
\frac{\dot u\,w-u\,\dot w}{w^2}
=
\alpha D^2+\beta D+\gamma.
\label{eq:appendix-projective-riccati-recovery}
\end{equation}

Thus the nonlinear Riccati equation is exactly the projective image of
the two-dimensional linear system.

For real \(q\), the matrix \(A_R(q)\) has real entries and zero trace,
so \(A_R(q)\in\mathfrak{sl}(2,\mathbb R)\). Along a complex Mellin
contour, the same construction is understood in the complexified
algebra \(\mathfrak{sl}(2,\mathbb C)\).

Define the auxiliary linear transport matrix

\begin{equation}
M_R(\tau,q)
=
e^{\tau A_R(q)}
=
\begin{pmatrix}
m_{11}(\tau,q)&m_{12}(\tau,q)\\
m_{21}(\tau,q)&m_{22}(\tau,q)
\end{pmatrix}.
\label{eq:appendix-projective-MR}
\end{equation}

The initial condition \(D(0,q)=0\) may be represented by

\begin{equation*}
u(0,q)=0,
\qquad
w(0,q)=1.
\end{equation*}

Hence

\begin{equation*}
\begin{pmatrix}
u(\tau,q)\\
w(\tau,q)
\end{pmatrix}
=
M_R(\tau,q)
\begin{pmatrix}
0\\
1
\end{pmatrix}
=
\begin{pmatrix}
m_{12}(\tau,q)\\
m_{22}(\tau,q)
\end{pmatrix},
\end{equation*}

and therefore

\begin{equation}
D(\tau,q)
=
\frac{m_{12}(\tau,q)}{m_{22}(\tau,q)}.
\label{eq:appendix-D-projective-ratio}
\end{equation}

The projective representation is valid as long as
\(m_{22}(s,q)\neq0\) for \(0\leq s\leq\tau\). A zero of
\(m_{22}\) corresponds to a pole of the Riccati solution. Along a
complex Mellin contour these poles must therefore be avoided; for real
transform arguments they delimit the corresponding exponential-moment
domain.

To evaluate \(M_R\) explicitly, note that

\begin{equation*}
A_R(q)^2=\Delta_q^2 I,
\qquad
\Delta_q^2=\frac{\beta^2}{4}-\alpha\gamma.
\end{equation*}

The exponential can therefore be evaluated directly:

\begin{equation}
M_R(\tau,q)
=
\cosh(\Delta_q\tau)I
+
\frac{\sinh(\Delta_q\tau)}{\Delta_q}A_R(q).
\label{eq:appendix-projective-exponential}
\end{equation}

At \(\Delta_q=0\), the quotient
\(\sinh(\Delta_q\tau)/\Delta_q\) is understood by continuity, with
limiting value \(\tau\).

For convenience, define

\begin{equation*}
B_\tau(q):=m_{12}(\tau,q),
\qquad
E_\tau(q):=m_{22}(\tau,q).
\end{equation*}

Equation~\eqref{eq:appendix-projective-exponential} gives

\begin{equation}
B_\tau(q)
=
\frac{\gamma}{\Delta_q}\sinh(\Delta_q\tau),
\qquad
E_\tau(q)
=
\cosh(\Delta_q\tau)
-
\frac{\beta}{2\Delta_q}\sinh(\Delta_q\tau).
\label{eq:appendix-projective-BE}
\end{equation}

Consequently,

\begin{equation}
D(\tau,q)
=
\frac{B_\tau(q)}{E_\tau(q)}
=
\frac{\gamma\sinh(\Delta_q\tau)}
{\Delta_q\cosh(\Delta_q\tau)-\frac12\beta\sinh(\Delta_q\tau)}.
\label{eq:appendix-D-explicit}
\end{equation}

The same projective construction also gives the amplitude factor entering the
momentum propagator, equation \eqref{eq:heston-affine-amplitude-mellin}:

\begin{equation*}
\frac{dA}{d\tau}
=
-(r-\delta)q+\kappa\theta D(\tau,q)-r,
\qquad
A(0,q)=0.
\end{equation*}

Thus

\begin{equation}
A(\tau,q)
=
-\bigl[(r-\delta)q+r\bigr]\tau
+
\kappa\theta\int_0^\tau D(s,q)\,ds.
\label{eq:appendix-A-integral}
\end{equation}

For \(\alpha\neq0\), the integral of \(D\) follows directly from the
equation for \(w\). Since \(w=E_\tau\) and

\begin{equation*}
\frac{\dot w}{w}
=
-\alpha D-\frac12\beta,
\end{equation*}

we obtain

\begin{equation*}
D
=
-\frac{1}{\alpha}
\left(
\frac{d}{d\tau}\log E_\tau+\frac12\beta
\right).
\end{equation*}

Integrating from \(0\) to \(\tau\), and using \(E_0=1\), gives

\begin{equation}
\int_0^\tau D(s,q)\,ds
=
-\frac{1}{\alpha}
\left[
\log E_\tau(q)+\frac12\beta\tau
\right].
\label{eq:appendix-D-integral}
\end{equation}

Substitution into \eqref{eq:appendix-A-integral} yields

\begin{equation}
A(\tau,q)
=
-\bigl[(r-\delta)q+r\bigr]\tau
-
\frac{\kappa\theta}{\alpha}
\left[
\log E_\tau(q)+\frac12\beta\tau
\right].
\label{eq:appendix-A-log}
\end{equation}

The case \(\alpha=0\) must be treated separately because
\eqref{eq:appendix-D-integral} contains \(1/\alpha\). The Riccati
equation then becomes linear,

\begin{equation*}
\frac{dD}{d\tau}=\beta D+\gamma,
\qquad
D(0,q)=0.
\end{equation*}

For \(\beta\neq0\),

\begin{equation}
D(\tau,q)
=
\frac{\gamma}{\beta}
\left(e^{\beta\tau}-1\right),
\qquad
\int_0^\tau D(s,q)\,ds
=
\frac{\gamma}{\beta^2}
\left(e^{\beta\tau}-1-\beta\tau\right).
\label{eq:appendix-linear-riccati}
\end{equation}

If \(\alpha=\beta=0\),

\begin{equation*}
D(\tau,q)=\gamma\tau,
\qquad
\int_0^\tau D(s,q)\,ds=\frac12\gamma\tau^2.
\end{equation*}

Along the complex Mellin contour, \(E_\tau(q)\) is complex and the
logarithm in \eqref{eq:appendix-A-log} must be evaluated on a branch
that is continuous along the contour and along the evolution in
\(\tau\). The contour must also avoid zeros of \(E_\tau(q)\), which
correspond to poles of the Riccati characteristic. These conditions
prevent the branch discontinuities that arise in direct numerical
implementations of the Heston characteristic function
\cite{KahlJaeckel2005,LordKahl2010}.


\appsection{Invariant Fields and Cartan Geometry}
\label{app:invariant-cartan-geometry}

This appendix collects derivations of the finite-sector central extension,
the corresponding left-invariant fields, the Poincar\'e--Cartan form, and
the characteristic field used in the main text. The finite Lie-group
construction is determined by the homogeneous quadratic sector \(C_s\).
The complementary sector \(C_H\) is not incorporated into the finite
group multiplication; it enters separately through multiplicative
holonomy and contributes to the full Heston Poincar\'e--Cartan form.

\subsection{Associativity of the Finite Central Extension}
\label{app:central-extension-associativity}

Let

\begin{equation*}
g=(t,\mathbf a),
\qquad
g'=(t',\mathbf a'),
\qquad
g''=(t'',\mathbf a''),
\qquad
\mathbf a=(\mathbf x,\mathbf p)^T=(x,v,p_x,p_v)^T.
\end{equation*}

The finite symplectic sector is the semidirect product

\begin{equation}
G^s
=
\mathbb R\ltimes_{M_s}\mathbb R^4,
\qquad
g'\star^s g
=
\left(
t'+t,
M_s(t)\mathbf a'+\mathbf a
\right).
\label{eq:app-finite-group-law}
\end{equation}

Here

\begin{equation*}
M_s(t)=e^{tK_s},
\qquad
M_s(t)^T J_4 M_s(t)=J_4.
\end{equation*}

The multiplicative \(\mathbb R_+\)-central extension
\(\widetilde G^s=G^s\times\mathbb R_+\) is defined by

\begin{equation}
(g',\zeta')
\widetilde\star^s
(g,\zeta)
=
\left(
g'\star^s g,
\zeta'\zeta
\exp\!\left[\epsilon_2^s(g',g)\right]
\right),
\qquad
\epsilon_2^s(g',g)
=
\frac12
\left(M_s(t)\mathbf a'\right)^T
J_4\mathbf a.
\label{eq:app-central-extension-law}
\end{equation}

Associativity of the lifted multiplication is equivalent to the
two-cocycle identity

\begin{equation}
\epsilon_2^s(g'',g')
+
\epsilon_2^s(g''\star^s g',g)
=
\epsilon_2^s(g',g)
+
\epsilon_2^s(g'',g'\star^s g).
\label{eq:app-cocycle-identity}
\end{equation}

Using \eqref{eq:app-finite-group-law}, the left-hand side of
\eqref{eq:app-cocycle-identity} is

\begin{equation*}
\frac12
\left(M_s(t')\mathbf a''\right)^T
J_4\mathbf a'
+
\frac12
\left[
M_s(t)
\left(
M_s(t')\mathbf a''+\mathbf a'
\right)
\right]^T
J_4\mathbf a.
\end{equation*}

Expanding the second term gives

\begin{equation*}
\frac12
\left(M_s(t')\mathbf a''\right)^T
J_4\mathbf a'
+
\frac12
\left(M_s(t+t')\mathbf a''\right)^T
J_4\mathbf a
+
\frac12
\left(M_s(t)\mathbf a'\right)^T
J_4\mathbf a.
\end{equation*}

The right-hand side is

\begin{equation*}
\frac12
\left(M_s(t)\mathbf a'\right)^T
J_4\mathbf a
+
\frac12
\left(M_s(t+t')\mathbf a''\right)^T
J_4
\left(
M_s(t)\mathbf a'+\mathbf a
\right).
\end{equation*}

Expanding the last term gives

\begin{equation*}
\frac12
\left(M_s(t)\mathbf a'\right)^T
J_4\mathbf a
+
\frac12
\left(M_s(t+t')\mathbf a''\right)^T
J_4M_s(t)\mathbf a'
+
\frac12
\left(M_s(t+t')\mathbf a''\right)^T
J_4\mathbf a.
\end{equation*}

Since

\begin{equation*}
M_s(t+t')=M_s(t)M_s(t')
\end{equation*}

and \(M_s(t)\) is symplectic,

\begin{equation}
\left(M_s(t+t')\mathbf a''\right)^T
J_4M_s(t)\mathbf a'
=
\left(M_s(t')\mathbf a''\right)^T
J_4\mathbf a'.
\label{eq:app-cocycle-symplectic-identity}
\end{equation}

The two sides of \eqref{eq:app-cocycle-identity} therefore agree.
Hence \(\epsilon_2^s\) is a two-cocycle and the lifted multiplication
on \(\widetilde G^s\) is associative.

The holonomy sector is separate from this finite central extension.


\subsection{Derivation of the Left-Invariant Fields of the Finite Symplectic Sector}
\label{app:left-invariant-fields}

The left-invariant fields of \(\widetilde G^s\) are obtained by
differentiating the group product with respect to the second factor at the
identity.
Introduce the positive central generator
\begin{equation}
\Xi=\zeta\partial_\zeta.
\label{eq:app-central-generator}
\end{equation}

Consider first an infinitesimal displacement in the phase-space direction
\(\boldsymbol\eta\in\mathbb R^4\),
\begin{equation*}
h_\epsilon=(0,\epsilon\boldsymbol\eta,1).
\end{equation*}
From the lifted group law,
\begin{equation}
(t,\mathbf a,\zeta)\widetilde\star^s h_\epsilon
=
\left(
t,\mathbf a+\epsilon\boldsymbol\eta,
\zeta\exp\!\left[
\frac{\epsilon}{2}\mathbf a^TJ_4\boldsymbol\eta
\right]
\right).
\label{eq:app-left-phase-infinitesimal-product}
\end{equation}
Differentiating at \(\epsilon=0\) gives
\begin{equation}
L_{\boldsymbol\eta}^s
=
\boldsymbol\eta^T
\left(
\nabla_{\mathbf a}-\frac12J_4\mathbf a\,\Xi
\right),
\qquad
\boldsymbol L_{\mathbf a}^s
=
\nabla_{\mathbf a}-\frac12J_4\mathbf a\,\Xi.
\label{eq:app-left-phase-field}
\end{equation}

For an infinitesimal time displacement
\begin{equation*}
h_\epsilon=(\epsilon,\mathbf 0,1),
\end{equation*}
one obtains
\begin{equation}
(t,\mathbf a,\zeta)\widetilde\star^s h_\epsilon
=
\left(t+\epsilon,M_s(\epsilon)\mathbf a,\zeta\right).
\label{eq:app-left-time-infinitesimal-product}
\end{equation}
Since \(M_s(\epsilon)=I+\epsilon K_s+o(\epsilon)\), differentiation gives
\begin{equation}
L_t^s=\partial_t+(K_s\mathbf a)^T\nabla_{\mathbf a},
\qquad
L_\zeta^s=\Xi.
\label{eq:app-left-time-field}
\end{equation}

With
\begin{equation*}
K_s=
\begin{pmatrix}
-\mathsf B&0\\
0&\mathsf B^T
\end{pmatrix},
\end{equation*}
the phase-space form becomes
\begin{equation}
L_t^s
=
\partial_t-(\mathsf B\mathbf x)^T\nabla_{\mathbf x}
+(\mathsf B^T\mathbf p)^T\nabla_{\mathbf p},
\qquad
\boldsymbol L_{\mathbf x}^s
=
\nabla_{\mathbf x}-\frac12\mathbf p\,\Xi,
\qquad
\boldsymbol L_{\mathbf p}^s
=
\nabla_{\mathbf p}+\frac12\mathbf x\,\Xi.
\label{eq:app-left-fields-phase-space}
\end{equation}


\subsection{Derivation of the Heston Poincar\'e--Cartan Form}
\label{app:cartan-form-derivation}

\subsubsection{Cartan Form of the Finite Symplectic Sector}

We first derive the Cartan form of the finite symplectic sector. It is the
left-invariant one-form \(\Theta_s\) normalized by
\begin{equation}
\Theta_s(\Xi)=1,\qquad
\Theta_s(\boldsymbol L_{\mathbf a}^s)=0,\qquad
\Theta_s(L_t^s)=0.
\label{eq:app-cartan-duality}
\end{equation}

Write
\begin{equation}
\Theta_s
=
\frac{d\zeta}{\zeta}
+\boldsymbol\alpha^Td\mathbf a+\alpha_t\,dt.
\label{eq:app-cartan-ansatz}
\end{equation}
The condition
\(\Theta_s(\boldsymbol L_{\mathbf a}^s)=0\) gives
\begin{equation}
\boldsymbol\alpha^T=-\frac12\mathbf a^TJ_4.
\label{eq:app-cartan-alpha}
\end{equation}

The quadratic sector satisfies
\begin{equation}
C_s=\frac12\mathbf a^TJ_4K_s\mathbf a.
\label{eq:app-Cs-quadratic}
\end{equation}
Therefore the condition \(\Theta_s(L_t^s)=0\) yields
\begin{equation}
\alpha_t=C_s.
\label{eq:app-cartan-alpha-time}
\end{equation}

Hence
\begin{equation}
\Theta_s
=
\frac{d\zeta}{\zeta}
-\frac12\mathbf a^TJ_4\,d\mathbf a
+C_s\,dt.
\label{eq:app-finite-cartan-form}
\end{equation}
Equivalently, in state and momentum variables,
\begin{equation}
\Theta_s
=
\frac{d\zeta}{\zeta}
+\frac12\mathbf p^Td\mathbf x
-\frac12\mathbf x^Td\mathbf p
+C_s\,dt.
\label{eq:app-finite-cartan-xp}
\end{equation}

\subsubsection{Poincar\'e--Cartan Form}

For an infinitesimal time displacement, the holonomy of the complementary
sector satisfies
\begin{equation*}
\mathcal H=1-C_H\,dt+o(dt).
\end{equation*}
Its infinitesimal action on the positive central fiber therefore contributes
the vertical term \(-C_H\Xi\), so the affine time field is
\begin{equation}
L_t^A=L_t^s-C_H\Xi.
\label{eq:app-affine-time-field}
\end{equation}

The full Poincar\'e--Cartan form is therefore
\begin{equation}
\Theta
=
\Theta_s+C_H\,dt
=
\frac{d\zeta}{\zeta}
-\frac12\mathbf a^TJ_4\,d\mathbf a
+C_A\,dt,
\qquad
C_A=C_s+C_H.
\label{eq:app-full-poincare-cartan}
\end{equation}
Equivalently,
\begin{equation}
\Theta
=
\frac{d\zeta}{\zeta}
+\frac12\mathbf p^Td\mathbf x
-\frac12\mathbf x^Td\mathbf p
+C_A\,dt.
\label{eq:app-full-poincare-cartan-xp}
\end{equation}

Its curvature is
\begin{equation}
\omega=d\Theta
=
-\frac12d\mathbf a^TJ_4\wedge d\mathbf a+dC_A\wedge dt
=
-d\mathbf x^T\wedge d\mathbf p+dC_A\wedge dt.
\label{eq:app-full-curvature}
\end{equation}

Thus \(C_s\) determines the finite-sector Cartan geometry, while the
holonomy contribution \(C_H\) converts \(\Theta_s\) into the full Heston
Poincar\'e--Cartan form.


\subsection{Derivation of the Characteristic Module}
\label{app:characteristic-module}

The characteristic module of the full Poincar\'e--Cartan form is the set of fields determined by the conditions
\begin{equation}
\mathcal C_\Theta
=
\left\{
X:\Theta(X)=0,\qquad \omega(X,\cdot)=0
\right\}.
\label{eq:app-characteristic-module-definition}
\end{equation}

For the Heston Poincar\'e--Cartan form
\eqref{eq:app-full-poincare-cartan-xp}, the characteristic module is locally of rank
one. We normalize its generator by \(dt(X_\Theta)=1\) and write
\begin{equation}
X_\Theta
=
\partial_t+\mathbf u^T\nabla_{\mathbf x}
+\mathbf w^T\nabla_{\mathbf p}+\eta\,\Xi.
\label{eq:app-characteristic-ansatz}
\end{equation}

Using
\begin{equation*}
\omega=-d\mathbf x^T\wedge d\mathbf p+dC_A\wedge dt,
\end{equation*}
the condition \(\omega(X_\Theta,\cdot)=0\) gives
\begin{equation}
\mathbf w=\nabla_{\mathbf x}C_A,\qquad
\mathbf u=-\nabla_{\mathbf p}C_A.
\label{eq:app-characteristic-spatial-components}
\end{equation}

Since
\begin{equation*}
C_A=F(\mathbf p)+\mathbf x^T\mathbf R(\mathbf p),
\end{equation*}
we have
\begin{equation}
\nabla_{\mathbf x}C_A=\mathbf R(\mathbf p).
\label{eq:app-characteristic-R}
\end{equation}

The condition \(\Theta(X_\Theta)=0\) determines the central coefficient:
\begin{equation}
\eta
=
\frac12
\left(
\mathbf p^T\nabla_{\mathbf p}
+\mathbf x^T\nabla_{\mathbf x}
-2
\right)C_A.
\label{eq:app-characteristic-eta}
\end{equation}

Hence
\begin{equation}
X_\Theta
=
\partial_t
+\mathbf R(\mathbf p)^T\nabla_{\mathbf p}
-\left(\nabla_{\mathbf p}C_A\right)^T\nabla_{\mathbf x}
+\eta\,\Xi.
\label{eq:app-characteristic-field}
\end{equation}

In four-dimensional symplectic notation,
\begin{equation}
X_\Theta
=
\partial_t
-\left(J_4\nabla_{\mathbf a}C_A\right)^T\nabla_{\mathbf a}
+\eta\,\Xi.
\label{eq:app-characteristic-field-4d}
\end{equation}

For Heston,
\begin{equation*}
\mathbf R(\mathbf p)
=
\begin{pmatrix}0\\ R(p_x,p_v)\end{pmatrix},
\qquad
R(p_x,p_v)
=
\frac12(p_x^2-p_x)
+(\rho\sigma_\nu p_x-\kappa)p_v
+\frac12\sigma_\nu^2p_v^2.
\end{equation*}

The explicit characteristic field is
\begin{equation}
\begin{aligned}
X_\Theta
={}&
\partial_t
+
\left[
\frac12(p_x^2-p_x)
+(\rho\sigma_\nu p_x-\kappa)p_v
+\frac12\sigma_\nu^2p_v^2
\right]\partial_{p_v}\\
&-
\left[
(r-\delta)
+v\left(p_x-\frac12+\rho\sigma_\nu p_v\right)
\right]\partial_x
-
\left[
\kappa\theta
+v\left(\rho\sigma_\nu p_x+\sigma_\nu^2p_v-\kappa\right)
\right]\partial_v
+\eta\,\Xi.
\end{aligned}
\label{eq:app-characteristic-field-heston}
\end{equation}

The central coefficient is
\begin{equation}
\eta
=
r-\frac12\left[(r-\delta)p_x+\kappa\theta p_v\right]
+\frac14v\left(
p_x^2+2\rho\sigma_\nu p_xp_v+\sigma_\nu^2p_v^2
\right).
\label{eq:app-characteristic-eta-heston}
\end{equation}

In terms of the left-invariant fields of the finite symplectic sector, the
same characteristic field can be written as
\begin{equation}
X_\Theta
=
L_t^s
-\left(\nabla_{\mathbf p}C_H\right)^T\boldsymbol L_{\mathbf x}^s
+\left(\nabla_{\mathbf x}C_H\right)^T\boldsymbol L_{\mathbf p}^s
-C_H\Xi.
\label{eq:app-characteristic-left-frame}
\end{equation}

In this representation, the contribution of \(C_s\) is already contained
in \(L_t^s\), while the complementary sector \(C_H\) appears explicitly.

\section*{Acknowledgments}
I thank the late Professor Peter Carr for helpful advice and valuable insights
on earlier ideas related to this work. I am also grateful to Gregory Pelts for
sharing his knowledge of the geometric structure underlying finance.
\vskip 3mm

\section*{Disclaimer}
\vskip 3mm
The views expressed herein are those of the author and do not reflect the views
of my current employer, Wells Fargo Securities, or affiliated entities.

\clearpage
\phantomsection

\end{document}